\documentclass{ws-ijmpa}
\usepackage{graphicx}
\usepackage{graphics}
\usepackage{amssymb}
\usepackage{amsmath}
\usepackage{amsfonts}
\usepackage[british]{babel}
\usepackage{hhline}
\usepackage{multirow}

\def\ra{{\rightarrow}}
\def\nn{{\nonumber}}
\def\mc{\mathcal}

\def\lm{{\lambda}}
\def\tta{{\theta}}
\def\om{{\omega}}
\def\ph{{\phi}}
\def\s{{\sigma}}
\def\al{{\alpha}}
\def\dl{{\delta}}
\begin{document}
\markboth{H. Saveetha, D. Indumathi }
{Fragmentation of $\omega$ and $\phi$ Mesons in $e^+\,e^-$ and 
$p\,p$ Collisions at NLO}

%
\catchline{}{}{}{}{}
%

\title{Fragmentation of $\omega$ and $\phi$ Mesons in $e^+\,e^-$ and $p\,p$ Collisions at NLO}

\author{H. Saveetha}
\address{Institute of Mathematical Sciences, \\ CIT Campus, \\ Chennai 600 113, India \\ saveetha@imsc.res.in}

\author{D. Indumathi}
\address{Institute of Mathematical Sciences, \\ CIT Campus, \\ Chennai 600 113, India \\ indu@imsc.res.in}

\begin{history}
\received{Day Month Year}
\revised{Day Month Year}
\end{history}

\begin{abstract}

A combined analysis of both $e^+\,e^-$ (LEP, SLD) and $p\,p$ (RHIC-PHENIX
and LHC-ALICE) hadroproduction processes are done for the first time for
the vector meson nonet at the next-to-leading order (NLO) using a model
with broken SU(3) symmetry. The transverse momentum ($p_T$) and rapidity
($y$) dependence of the differential cross section for $\omega$ and $\phi$
mesons of the $p\,p$ data are also discussed.  The input universal quark
(valence and singlet) fragmentation functions at a starting scale of
$Q_0^2=1.5$ GeV$^2$, after evolution, have values that are consistent
with the earlier analysis for $e^+\,e^-$ at NLO. However, the universal
gluon fragmentation function is now well determined from this study with
significantly smaller error bars, as the $p\,p$ hadroproduction cross
section is particularly sensitive to the gluon fragmentation since it
occurs at the same order as quark fragmentation, in contrast to the
$e^+\,e^-$ hadroproduction process. Additional parameters involved in
describing strangeness and sea suppression and octet-singlet mixing
are found to be close to earlier analysis; in addition, a new relation
between gluon and sea suppression in $K^*$ and $\phi$ hadroproduction
has been observed.


\end{abstract}

\ccode{PACS numbers: 13.60.Le, 13.60.Hb, 13.66.Bc, 13.85.Ni}

\maketitle

\section{Introduction}
A number of analyses are available for fragmentation of
pseudoscalar mesons and baryons till date; see, for example,
Ref.\cite{Kret} for $\pi$, 
Refs.\cite{Indu,Misra,Albino1,Albino2,Albino3,Kumano1,Kumano2,Strat} 
for $K$ meson,
Refs.\cite{Hirai,Flor} for proton, and Refs.\cite{Misra,Stratmann}
for $\eta$ fragmentation with comprehensive reviews in
Refs.\cite{Albino1,Albino2,Albino3,Ams} as well. No such considerable
interest has been shown towards vector meson production due to the
scarcity of the data available so far.

Hadroproduction of $\phi$ vector mesons in proton-proton collisions
is a good candidate signal for Quark Gluon Plasma (QGP) in heavy
nucleus-nucleus collisions. This requires a good understanding of $\phi$
hadroproduction in $p\,p$ collisions, which will serve as a baseline
for nucleus--nucleus studies. With this motivation, analyses had been
done for vector meson fragmentation in $e^+\,e^-$ scattering at Leading
Order (LO)\cite{Savilo} and Next-to-Leading Order (NLO)\cite{Savinlo}
and in $p\,p$ collisions at LO\cite{Savilo} as well. In this paper,
for the first time, this has been extended to a combined investigation
for the fragmentation of the entire vector meson nonet in $e^+\,e^-$ and
for $\om$ and $\phi$ meson production in $p\,p$ collisions at NLO using
the LEP\cite{Data,Rho1,Rho2,Rho3,Rho5,Omega1,Omega2} and
SLD\cite{SLD1,SLD2} data for $e^+\,e^-$
and RHIC PHENIX\cite{RHIC} data and LHC ALICE\cite{LHC-17} data for
$p\,p$ hadroproduction. Analysis of individual vector meson production
has been done, for instance, analysis of $\phi$ hadroproduction from
LHC data was done in Ref.~\cite{Wei}.

A key feature of the analysis (described earlier in Refs.\cite{Savilo}
and \cite{Savinlo} and applied to the present study) is the ability
to use the entire nonet vector meson hadroproduction data by defining
SU(3)-symmetric fragmentation functions common to the entire set of octet
mesons. This drastically reduces the number of independent fragmentation
functions (from three quark- and one gluon fragmentation function for
{\em each} member of the octet) to two universal quark- and one gluon
fragmentation function. Some additional parameters are subsequently
introduced to account for SU(3) symmetry breaking and singlet-octet
mixing to allow the study of the entire vector meson nonet. The definition of
input fragmentation functions and other parameters relevant to the
model\cite{Savinlo} remain the same in this study and have been briefly
reviewed here for completeness; some differences in the choice of
fragmentation functions for analyses are also mentioned below. The
hadroproduction in $p\,p$ collisions at NLO presented here is particularly
important in view of the fact that gluons contribute at higher order in
$e^+\,e^-$ collisions but contribute at the same order as quarks in $p\,p$
processes. This study at NLO therefore enables a more precise determination of
the gluon fragmentation functions.

Further studies such as gluon and singlet quark suppression, the
dependence of the $p\,p$ hadroproduction cross sections on transverse
momentum $p_T$ as well as rapidity $y$, and inclusion of data on
the (branching fraction-weighted) $\phi$ and $\omega$ cross section
ratios helps in a more detailed understanding of the hadroproduction
process. With this study we complete the program of vector meson nonet
fragmentation using this model.

In section \ref{Kin}, we list the relevant cross-section formulae for
hadroproduction in $e^+\,e^-$ and $p\,p$ collisions. In Section \ref{Mod}, 
we
present highlights of the model used to determine the vector meson
fragmentation functions. In Section \ref{Fit} we use the available data to
best-fit the parameters involved and show the resultant fits and their
quality. We conclude in Section \ref{Sum} with some remarks and summary.

\section{Kinematics and Cross sections}
\label{Kin}
We summarise here for completeness the relevant cross-sections for
inclusive hadroproduction in $e^+\,e^-$ and $p\,p$ collisions (in the
c.m. frame).

\subsection{Hadroproduction in $e^+\,e^-$ Collisions}
\label{EEprod}
The hadronic cross section for inclusive hadroproduction in $e^+\,e^-$
collisions to NLO is given by\cite{Fur}:
\begin{eqnarray} \nonumber
\frac{1}{\s_{tot}} \frac{d\s^{h}_{e^+e^-}}{dx}(x;Q^2)
& = & \frac{1}{\sum_F\lm_B^F \left(1 + {\al_s}/{\pi}\right)}
\int_x^1 \frac{dz}{z}\left[ \sum_F
\lm_B^F\left(\dl(1-z)+ \right. \right. \\ \nonumber
 & & \left. \left. \frac{\al_s(Q^2)}{2\pi} \mc C^{F(1)}(z) \right)
\left\{D^h_{q_f} +D^h_{\bar q_f}\right\}\left(\frac
xz\right) + \frac{\al_s(Q^2)}{2\pi}\lm_B^g\mc
        C^{g(1)}(z)D^h_g\left(\frac xz\right)\right]~, \nn \\
\hbox{and  }
\s_{tot} & = & N_c \sum_F{\lm_B^F}{\left(\frac{4\pi\al^2}{3Q^2}\right)}
\left( 1 + \frac{\al_s}{\pi}\right)~.
\label{eq:coeff}
\end{eqnarray}
Here $x (=2 p_h/\sqrt{s})$ is the fraction of the parent quark momentum
carried by the hadron $(h)$ having momentum $p_h$, $Q= \sqrt{s}$ is the
energy scale at which the analysis is carried out (the data is taken
at the $Z$-pole, with $Q = 91.2$ GeV), functions like $D^h_{q_f}$,
$D^h_{\bar{q_{f}}}$ and $D^h_g$ are the quark, anti-quark and gluon
fragmentation functions and $N_c$ refers to the number of colours.
Terms like $\mc C^{F(1)}(z)$, $\mc C^{g(1)}(z)$, $\lm_B^F$ and $\lm_B^g$
are the coefficient functions for quarks (F) and gluons (g), whose
expressions are given in detail in Refs.\cite{Savinlo,Fur}
and where $\alpha_s (Q^2)$ is also defined to NLO.

\subsection{Hadroproduction in $p\,p$ Collisions}
\label{PPprod}

The hadronic cross section for inclusive hadroproduction in $p\,p$ collisions
at NLO is given in terms of the underlying partonic interaction $p_i(x_1) \,
p_j (x_2) \rightarrow p_l(x_3)\,  p_k(x_4)$ as\cite{Aversa},
\begin{eqnarray}
E_3 \frac{d^3\sigma}{d^3k_3} &\sim& \sum_{i,j,l} \int dx_1dx_2
       \frac{dz}{z^2} f_{p_i/H_1}(x_1, M^2)
        f_{p_j/H_2}(x_2, M^2) D_{p_l/H_3}(z, M^2_f) \nn \\
 & & \times \left[\frac{1}{v} \left( \frac{d\sigma^0}{dv}
       \right)_{p_ip_j \ra p_l} (s,v) \delta(1-w) +
       \frac{\alpha_s(\mu^2)}{2\pi}K_{p_ip_j \ra p_l}
       (s,v,w;\mu^2;M^2,M^2_f)\right]~,
\label{eq:crosssec}
\end{eqnarray}
where the indices $i, j, l,$ sum over all possible flavours of
quarks and anti quarks, and gluons. The term $f_{p_i/H_1}(x_1, M^2)$
($f_{p_j/H_2}(x_2, M^2)$) refers to the parton distribution function
of parton $p_i ( p_j)$ inside hadron $H_1 (H_2)$ with a momentum
fraction $x_1 (x_2)$ and initial factorization scale $M$. Likewise,
$D_{p_l/H_3}(z, M^2_f)$ is the fragmentation function for a parton
$p_l$ to fragment into a hadron $H_3$ with a momentum fraction $z$
and fragmentation scale $M_f$.

The first term within the bracket, $d\sigma^0$, is the LO Born cross
section term for $p_i p_j \ra p_l$ with $s, v$ and $w$ expressed in terms
of $x_1, x_2$ and $z$ and hadronic momenta $k's$; for example, $s = x_1
x_2 S$ where $S$ is the usual hadronic centre of mass energy (squared).

The second term having $\alpha_s(\mu^2)$ with renormalization scale
$\mu$ corresponds to the higher order contribution with its correction
factor $K_{p_i p_j \ra p_l}(s,v,w;\mu^2;M^2,M^2_f)$ for each subprocess.
A detailed calculation of the correction factors for various subprocesses
is given in Ref.\cite{Aversa}; here we merely note that, unlike in the
$e^+\,e^-$ case, in $p\,p$ processes the gluon fragmentation function
contributes at the LO itself through subprocesses such as $q \, g \to q \,
g$ and $g \, g \to g \, g$. Hence we expect that inclusion of
hadroproduction in $p\,p$ processes will significantly improve our
knowledge of gluon fragmentation.

Hence from Eqs.~\ref{eq:coeff} and \ref{eq:crosssec} it is very
clear that the gluon fragmentation function appears at a higher order of
$\al(Q^2)$ as compared to quark fragmentation  in $e^+\,e^-$ processes
and at the same order in $p\,p$ processes.

The LHS of Eq.~\ref{eq:crosssec} can be expressed in terms of physical
observables, the rapidity $y$ and the transverse momentum $p_T$, as
\begin{equation}
E_3 \frac{d^3\sigma}{d^3k_3}  \equiv 
\frac{1}{p_T} \frac{d^3\sigma}{dp_T dy d\phi} = \frac{1}{\pi}
\frac{d^2\sigma}{dp_T^2 dy}~,
\label{eq:physobs}
\end{equation}
where the last simplification occurs because the cross-section is
independent of the azimuthal angle $\phi$.  According to the factorization
theorem, the cross section for $p\,p$ in Eq.~\ref{eq:crosssec} is
expressed as a convolution over three parts: parton distribution
functions, partonic subprocess cross sections and fragmentation
functions. For this study, the initial parton distribution functions
are taken from GRV 98 NLO code\cite{GRV98} (Within this limited $x$
region, the CTEQ parton distribution functions can be used as well),
the partonic cross sections for hadroproduction in $p\,p$ processes at
NLO are taken from Aversa et al.\cite{Aversacode}, and the fragmentation
of the final state parton is obtained using our $x$-real space Fortran
code based on the broken SU(3) model.

\section{The Model}
\label{Mod}
We now briefly describe the broken SU(3) model that is used to describe
the input fragmentation functions at NLO in this paper. The details
regarding the model were discussed in detail in Refs.\cite{Indu,Savilo,Savinlo} in which the $e^+\,e^-$ data were fitted
to the NLO cross sections using this model. In Ref.\cite{Savilo},
a study of hadroproduction in $p\,p$ processes at LO was also taken
up. The present study, which includes consistently an analysis of both
$e^+\,e^-$ and $p\,p$ hadroproduction to NLO completes this program.

The model uses SU(3) flavour symmetry to express the quark fragmentation
functions $\alpha(x, Q^2)$, $\beta(x, Q^2)$ and $\gamma(x, Q^2)$
corresponding to the underlying quark fragmentation subprocesses $q^i \to
M^i_j \, X^j$, where $X^j$ is a member of 3-, $\overline{6}$-, or 15-plet
respectively. Application of charge conjugation symmetry and isospin
invariance significantly reduces the number of unknown fragmentation
functions. In addition, fragmentation functions of different mesons are
related within this model, and this is what allows for the analysis of
the otherwise sparse vector meson data.

The fragmentation functions of all octet vector mesons can be written
in terms of three universal functions that are named valence $(V)$,
sea $(\gamma)$, and gluon $(D_g)$ fragmentation functions\cite{Savinlo}
(see Table~\ref{tab:frag}). The model defines the fragmentation functions
at an initial scale of $Q_0^2$, taken to be $Q_0^2 =1.5$ GeV$^2$, for
three light quarks $u$, $d$ and $s$, where the charm and bottom flavour
contributions are kept zero. The contribution of such heavy flavours are
added in at appropriate thresholds during DGLAP evolution. These input
fragmentation functions are then evolved to various momentum scales for
comparison with available data.

\begin{table}[tbh]
\tbl{Pure SU(3) quark fragmentation functions for octet mesons in
terms of the SU(3) valence ($V(x, Q^2)$) and sea ($\gamma (x, Q^2)$)
fragmentation functions. The valence quark content of the mesons is
indicated in brackets.}
{\begin{tabular}{ccl|ccl}
\toprule
fragmenting & \multicolumn{2}{c|}{${}_{\displaystyle K^{*+}}$} & fragmenting &
\multicolumn{2}{c}{${}_{\displaystyle K^{*0}}$} \\
quark & \multicolumn{2}{c|}{$(u \overline{s})$} & quark &
\multicolumn{2}{c}{$(d \overline{s})$} \\  \colrule
$u, \overline{s}$ & : &  $V + 2 \gamma$ &
$u, \overline{u}$ & : &  $2{\gamma}$ \\
$d, \overline{d}$ & : &  $2{\gamma}$ &
$d, \overline{s}$ & : &  $V + 2 {\gamma}$ \\
$s, \overline{u}$ & : &  $2 {\gamma}$ &
$s, \overline{d}$ & : &  $2 {\gamma}$ \\ \colrule
fragmenting & \multicolumn{2}{c|}{${}_{\displaystyle \omega}$} &
fragmenting &
\multicolumn{2}{c}{${}_{\displaystyle \rho^0}$} \\
quark & \multicolumn{2}{c|}{$((u \overline{u} + d \overline{d} - 2 s
\overline{s})/\sqrt{6})$}
& quark & \multicolumn{2}{c}{$((u \overline{u} - d
\overline{d})/\sqrt{2})$} \\  \colrule
$u, \overline{u}$ & : &
$\frac{1}{6}{V} + 2 {\gamma}$ &
$u, \overline{u}$ & : &
$\frac{1}{2}{V}+2 {\gamma}$ \\
$d, \overline{d}$ & : &
$\frac{1}{6}{V}+2 {\gamma}$ &
$d, \overline{d}$ & : &
$\frac{1}{2}{V}+2 {\gamma}$ \\
$s, \overline{s}$ & : &
$\frac{4}{6}{V}+2 {\gamma}$ &
$s, \overline{s}$ & : &  $2\gamma$ \\ \colrule
fragmenting & \multicolumn{2}{c|}{${{}_{\displaystyle \rho^+}}$} &
fragmenting &  \multicolumn{2}{c}{${{}_{\displaystyle \rho^-}}$} \\
quark & \multicolumn{2}{c|}{$(u \overline{d})$} & quark &
\multicolumn{2}{c}{$(d \overline{u})$} \\  \colrule
$u, \overline{d}$ &  : & $V + 2 {\gamma}$ &
$u, \overline{d}$ &  : & $2 {\gamma}$ \\
$d, \overline{u}$ &  : & $2 {\gamma}$ &
$d, \overline{u}$ &  : & $V + 2 {\gamma}$ \\
$s, \overline{s}$ &  : & $2 {\gamma}$ &
$s, \overline{s}$ &  : & $2 {\gamma}$ \\ \colrule
fragmenting & \multicolumn{2}{c|}{${{}_{\displaystyle \overline{K^{*0}}}}$} &
fragmenting & \multicolumn{2}{c}{${{}_{\displaystyle K^{*-}}}$} \\
quark & \multicolumn{2}{c|}{$(s \overline{d})$} & quark &
\multicolumn{2}{c}{$(s \overline{u})$} \\  \colrule
$u, \overline{u}$ & : & $2 {\gamma}$ &
$u, \overline{s}$ & : & $2 {\gamma}$ \\
$d, \overline{s}$ & : & $2 {\gamma}$ &
$d, \overline{d}$ & : & $2 {\gamma}$ \\
$s, \overline{d}$ & : & ${V} + 2 {\gamma}$ &
$s, \overline{u}$ & : & ${V} + 2 {\gamma}$ \\ \botrule
\end{tabular}
\label{tab:frag}}
\end{table}

Breaking of SU(3) symmetry due to strangeness suppression is included
through an $x$-independent strangeness suppression parameter $\lm$ at
the starting scale. For instance, non-strange quark fragmentation into
strange mesons such as $K^*$ is suppressed by $\lambda$: $D_u^{K*} \to
\lambda D_u^{K*, {\rm pure~SU(3)}}$ while strange quark fragmentation
is not suppressed: $D_s^{K*}$ = $D_s^{K*, {\rm pure~SU(3)}}$ (see
Table~\ref{tab:frag} for the pure SU(3) expressions). The entire sea quark
fragmentation into $K^*$ is thus suppressed by a factor of $\lambda$
compared to sea quark fragmentation into $\rho$ mesons.

The model is extended to include the SU(3) singlet-octet mixing since it
is known that the physical $\omega$ and $\phi$ mesons are admixtures of
the octet and singlet states. An angle $\theta$ is used to describe SU(3)
singlet--octet mixing. The singlet sector has an additional fragmentation
function, $\delta(x, Q^2)$, due to the single subprocess that contributes:
$q^i \to M^0 \, X^i$, where $X^i$ belongs to a 3-plet, which is taken
to be proportional to the octet fragmentation function $\alpha$:
$$
\frac{\delta(x, Q^2)}{3} = \frac{f_1^i}{3} \alpha(x, Q^2)~,
$$
thus adding only two parameters for $i = \omega, \phi$, viz., $f_1^{u,
\omega}$ and $f_1^{s, \phi}$. Note that
$f_1^{d,\omega}= f_1^{u,\omega}$ and
$f_1^{s,\omega}= f_1^{u,\phi} = f_1^{d,\phi} = 0$. The former arises from
SU(3) and SU(2) symmetry and the latter from the observation that the
physical $\phi$ ($\omega$) meson is almost purely an $s\overline{s}$
(non-strange) state since the phenomenological value of the mixing angle
$\theta \sim 39^\circ$ is very close to the value $\theta = 35^\circ$
where this is exactly true. Finally the sea suppression factors for
$\omega$ and $\phi$ are denoted as $f_{sea}^\omega$ and $f_{sea}^\phi$;
they are expected to be of order unity and $\lambda^2$ respectively. Note
that no additional {\emph {singlet fragmentation functions}} are required.

In toto, we have the fragmentation functions for octet valence, sea and
gluon ($V,\gamma$ and $D_g$) with strangeness suppression $\lm$, the
octet-singlet mixing angle $\tta$, and other $x$-independent singlet and 
suppression factors for the mixed $\omega$-$\phi$ system such as 
$f_1^u({\om})$, $f_1^s({\ph})$, $f_{sea}^{\om}$ and $f_{sea}^{\ph}$.  
Finally, we have the gluon suppression factors $f_g^{K^*}$, $f_g^{\om}$, 
$f_g^{\ph}$\cite{Savinlo}, where $D_g  = f_g^i D_g$, $i = K^*, \omega, \phi$.

The following modification has been made in the parameter descriptions compared to
the earlier analyses, leading to better stability during evolution:
We have used upto linear terms in $x$ instead of the choice of
a quadratic form in the standard polynomial\cite{Reya} for the
parameterization of input quark and gluon fragmentation functions:
\begin{equation}
F_i(x,Q_0^2) = a_i  x^{b_i}(1-x)^{c_i}\, P_i(x);
~~P_i(x) = (1 + d_i \sqrt x + e_i x)~;
\label{eq:func}
\end{equation}
instead of $P_i = (1 + d_i x + e_i x^2)$ which was the form used in the
earlier analysis. This polynomial can have large fluctuations and even
go negative during $Q^2$ evolution especially in the low-$x$ region,
while the current choice shows smooth behaviour at low and intermediate
values of $x$. Hence, this polynomial choice helps in obtaining a more
stable fit at low-$x$, and a better fit at intermediate-$x$, particularly
for $p\,p$ data.  Here $F_i(x) = V(x), \gamma(x)$ and $D_g(x)$ are the
corresponding valence, sea and gluon input fragmentation functions and
$a_i, b_i, c_i, d_i$ and $e_i$ are the parameters to be determined for
these functions at the starting scale $Q_0^2$.


\section{Combined Analysis of $e^+\,e^-$ and $p\,p$ Data }
\label{Fit}
\subsection{Choice of Data Sets}
\label{Choice}
A combined analysis of both $e^+\,e^-$ and $p\,p$ data is done in
order to fit the vector meson fragmentation functions.  The LEP
data\cite{Data,Rho1,Rho2,Rho3,Rho5,Omega1,Omega2} for
$\rho(\rho^+, \rho^-, \rho^0)$ and $\omega$
mesons and SLD ``pure uds'' data\cite{SLD1,SLD2} for K$^*$ and $\phi$ are used
for $e^+\,e^-$ process at the $Z$-pole, $\sqrt s = 91.2$ GeV. The SLD
``pure uds'' data (three flavours alone) are used in the case of K$^*$
and $\phi$ mesons in order to avoid the contamination from heavy flavour
meson production such as $B$ and $D$ mesons which decay into one of the
strange mesons which will contaminate the data on direct hadroproduction
into K$^*$ or $\phi$ due to large CKM matrix elements $V_{cs}$ and
$V_{cb}$. In the case of non-strange mesons like $\rho$ and $\omega$,
the contamination is very small, since heavier $b$- and $c$- mesons will
decay mostly (vis $s$) to $\pi$, the lightest non-strange pseudoscalar
meson, rather than $\rho$ or $\omega$.

Likewise, the 2011 RHIC/PHENIX data\cite{RHIC} at centre-of-mass energy,
$\sqrt{s} = 200$ GeV, with rapidity (to be considered as pseudorapidity
throughtout the paper), $|y|\le 0.35$ for
$p\,p$ collisions is used in the analysis for $\om$ and $\ph$ hadroproduction.
The data has three types of systematic errors added in quadrature
with no statistical errors given in the literature. Effort is taken to
add statistical errors from RHIC experimental group paper\cite{RHIC} and
thesis\cite{Deepali} for $\om$ and $\phi$ mesons decaying through various
channels. Thus care is taken to include both the statistical and systematical
errors which are added in quadrature. More recently\cite{Ratio},
detailed doubly differential rates in both rapidity $y$ and transverse
momentum $p_T$ have been measured by RHIC-PHENIX, in the forward
rapidity region $1.2 \le \vert y \vert \le 2.2$ for $\omega$ and $\phi$
hadroproduction, as well as their weighted events ratio. Recently, the
LHC-ALICE collaboration\cite{LHC-17} has also provided $K^*$ and $\phi$
hadroproduction data at $\sqrt{s} = 2.76$ TeV\footnote{We thank the
referee for bringing this to our notice.} Since the $K^*$ data is
sensitive to both the valence fragmentation function and the strangeness
suppression factor, $\lambda$, we have also included this data in our
analysis. 

The rapidity and azimuthal acceptances are different in different sets
of $p\,p$ data, and we have used the values to match with the
experimental data.

\subsection{Determining the Best-fit Parameters}
\label{Bestfit}
Using the standard functional form in Eq.~\ref{eq:func}, the unknown
input fragmentation functions for $V$, $\gamma$ and $D_g$ are parameterized
at an initial scale of $Q^2_0 = 1.5$ GeV$^2$.
Contributions
of the heavy $c$ and $b$ flavours are included at the appropriate thresholds
during evolution. The fragmentation functions of all mesons are
evolved to various scales, say, $Q^2 = (91.2)^2$ GeV$^2$ for $e^+\,e^-$ and
$Q^2 = p_T^2$ GeV$^2$ for $p\,p$ collision, using the
DGLAP evolution equations\cite{Dglap1,Dglap2,Dglap3} for $\rho$,
$K^*$, $\omega$ and $\phi$.

The best fit to the unknown parameters is found by performing a
combined $\chi^2$ minimization with both $e^+\,e^-$
LEP\cite{Data,Rho1,Rho2,Rho3,Rho5,Omega1,Omega2} and
SLD\cite{SLD1,SLD2} data and $p\,p$
RHIC-PHENIX (for both hadroproduction\cite{RHIC} and branching ratio weighted 
differential cross section\cite{Ratio}) and LHC-ALICE\cite{LHC-17} data.

The best fit values of the parameters $a, b, c, d$ and $e$
for valence, singlet and gluon input fragmentation functions, with
$1\sigma$ errors are given in Table~\ref{tab:inputs}. The errors on the
quark parameters are about 5\% or less, much better than the
earlier\cite{Savinlo}
studies. However, the fits to the gluon parameters are much
better determined than earlier, with errors of 1-4\% on the fit
parameters. This is due to two reasons, the first that the $p\,p$ cross
sections are sensitive to both quark and gluon fragmentation functions
at the same order, and the second that a huge energy sale separates RHIC
and LHC data, thereby restricting the allowed parameter space
considerably.

\begin{table}[tbh]
\tbl{Best fit values of the parameters defining the input octet
valence and sea quark and gluon fragmentation functions at the
starting scale of $Q_0^2 = 1.5$ GeV$^2$, with their $1$-$\sigma$
error bars.}
{\begin{tabular}{ccccc} \toprule
\multicolumn{2}{c}{parameter} & Central Value
& \multicolumn{2}{c}{Error Bars}   \\
\cline{4-5}
$V$     &$a$ & 2.42   & 0.30  &  0.29 \\
        &$b$ & 2.24   & 0.21  &  0.18 \\
        &$c$ & 2.71   & 0.13  &  0.12  \\
        &$d$ & 2.43   & 0.59  &  0.56 \\
        &$e$ & 1.17   & 0.78  &  0.74 \\ \hline
$\gamma$&$a$ & 0.32   & 0.01  &  0.02  \\
        &$b$ & -0.73  & 0.03  &  0.03  \\
        &$c$ & 3.53    & 0.13 &  0.12 \\
        &$d$ & 0.70   & 0.14  &  0.57 \\
        &$e$ & 0.42   & 0.26  &  0.26  \\ \hline
$D_g$   &$a$ & 2.43   & 0.07  &  0.07  \\
        &$b$ & 0.94   & 0.05  &  0.04 \\
        &$c$ & 2.68   & 0.03  &  0.03  \\
        &$d$ & -0.18  & 0.04  &  0.08  \\
        &$e$ & 1.04   & 0.07  &  0.07  \\ \botrule
\end{tabular}
\label{tab:inputs}}
\end{table}

\subsubsection{Quark fragmentation functions}
\label{Qffns}
While the behaviour of the valence quark fragmentation function is
similar to before, the small-$x$ behaviour is not as well determined.
This is expected since the small-$x$ is dominated by the sea quark
fragmentation functions. Still the parameters are much better
determined, the polynomial, $P_i(x)$, in particular,  is much better
determined---with smaller errors, falling from
nearly 100\% earlier to a few percent in the current analysis. These
parameters give the pure octet non-strange fragmentation functions for
$\rho^\pm, \rho^0$ mesons. The corresponding fragmentation functions for
$K^*$, $\omega$ and $\rho$ can be determined from these and the best-fit
values of the additional (strangeness suppression and singlet-octet
mixing) parameters that are listed in Table~\ref{tab:unknown_par}.

\begin{table}[tbh]
\tbl{Best fit values of the strangeness suppression factor
$\lambda$, the singlet--octet mixing angle, $\tta$, and other
suppression factors for $\omega$, $\phi$ hadroproduction and gluon
suppression factors at the initial scale of $Q_0^2 = 1.5$ GeV$^2$; note
that $f_{sea}^\phi$ has been set to $f_{sea}^\phi = \lm^2$. The positive
error bar on $f_{sea}^{\omega}$ is nominal, since the value cannot
exceed 1. For details see text.}
{\begin{tabular}{ccccc} \toprule
parameter        & Central Value   & \multicolumn{2}{c}{Error Bars} \\
                 &        & error  & error \\ \colrule
$\lm$            & 0.097  & 0.013  &  0.012\\
$\tta $          & 39.5  & 1.4   &  2.3 \\
$f_{sea}^{\om}$  & 0.99   & 0.10(*)   &  0.1  \\
$f_{sea}^{\phi}$  & $\lambda^2$   & const   &  --   \\
$f_1^u(\om)$     & 0.000  & --   &  0.08 \\
$f_1^s(\ph)$     & 7.48   & 1.75   &  1.61 \\
$f_g^{K^*}$      & 0.42   & 0.02   &  0.02 \\
$f_g^{\om}$      & 0.90   & 0.02   &  0.02 \\
$f_g^{\ph}$      & 0.22   & 0.01   &  0.01 \\ \botrule
\end{tabular}
\label{tab:unknown_par}}
\end{table}

The value of the strangeness suppression factor, $\lambda = 0.097 \pm
0.013$, is consistent with the previous analysis\cite{Savinlo} and also
with $\lambda^{pseudo}=  0.08$ for pseudoscalar mesons\cite{Misra} within
error bars. Hence it is very clear from the consistent value of $\lambda$,
that it is a process- and spin- independent global parameter.

The current sea fragmentation function 
at small $x$ is about $10$ times larger (with a large and negative exponent
$b$) due to the inclusion of low $p_T$ $p\,p$ data. While the $p_T$
range overlaps for the RHIC and LHC central rapidity data, the
corresponding $z$ ranges that they probe are entirely different, being
$0.01 \le z \le 1$ for RHIC and $0.001 \le z \le 1$ for LHC. Note that
a few LHC data points with $p_T < 2$ GeV were removed from the fits
since these corresponded to values of $z < 0.001$ where our fits are not
stable or reliable.
The small errors on the fits to the parameters describing $\gamma$ are
driven by this large range in $\sqrt{s}$
of the available $p\,p$ data. 
     
\subsubsection{Singlet-octet Mixing and $\omega$-$\phi$ fragmentation
functions}
\label{Mixing}
The best-fit value of the octet-singlet mixing angle, $\tta$, turns out
to be $\theta = 39.5^\circ \pm 1.3^\circ$ which is close to $35^\circ$
where $\om$ is a pure non-strange physical state\cite{Yao} and $\phi$
purely an $s\bar{s}$ state.  This angle is consistent with the value
$\tta=40.5^\circ$ determined from our earlier $e^+\,e^-$ NLO
studies\cite{Savinlo}.

The sea suppression factor, $f_{sea}^{\om}$ for $\om$ came out
to be $0.99 \pm 0.09$, the same as earlier, implying that there is no
suppression for sea in $\omega$ as it is purely nonstrange. The sea
suppression factor for $\phi$, $f_{sea}^{\phi}$, was kept fixed as before
to be equal to the square of the strangeness suppression ($\lambda ^2$)
since it is dominantly a pure $s \bar{s}$ state.

The singlet proportionality factors are obtained as $f_1^u(\om) = 0.00
\pm 0.08$ and $f_1^s(\ph)=7.48 \pm 1.7$ which are again consistent with
the earlier analysis.

\subsubsection{The gluon fragmentation functions}
\label{Gffns}
Through the $p\,p$ NLO fragmentation study, the gluon fragmentation
function and its suppression in strange mesons are better understood
compared to the earlier $e^+\,e^-$ analysis. The parameter values obtained
for input gluon fragmentation functions at the starting scale $Q_0^2$
(see Table~\ref{tab:inputs}) have very small error bars (much less than
$5\%$) compared to earlier studies\cite{Savilo,Savinlo}. For instance,
even the parameters $d$ and $e$ in the polynomial of Eq.~\ref{eq:func}
have error bars within $5\%$ which implies that the gluon parameters
are precisely determined through the analysis. The greatly reduced error
bars reflect the higher sensitivity of $p\,p$ hadroproduction to gluon
fragmentation.

Note that the $x$-dependence of the gluon fragmentation function at
small-$x$ is poorly determined from RHIC data but well constrained when
the LHC data is included. The small-$x$
behaviour of the gluon fragmentation function is similar to that
obtained earlier.

The addition of the $p\,p$ data allows a more precise determination of
the gluon fragmentation functions. Also, we have introduced three new
gluon suppression factors, $f_g^i$, $i= K^*,\om, \phi$ so that $D_g^i =
f_g^i D_g^{\rho}$. The gluon suppression factor for $\omega$ meson came
out to be $f_g^{\om} = 0.90 \pm 0.02$, as expected (and its value is
consistent with the one obtained in earlier analysis\cite{Savinlo}), but with
its error severely reduced due to the improved precision. Notice that
the current value is away from unity, and this is consistent with the
quark fragmentation functions also being marginally smaller than that
for $\rho$ hadroproduction. 

It turns out that the value of the gluon suppression factor for K$^*$
is $f_g^{K^{*}}=0.42 \pm 0.02$ which is quite different from
before\cite{Savinlo} ($1.0 \pm 0.09$).  Also, the value of gluon suppression
factor for $\phi$ meson obtained from the analysis $f_g^{\phi} = 0.22 \pm
0.01$ is very stable throughout this analysis; it is lower than the value
obtained from $e^+\,e^-$ case $(0.4\pm 0.04)$, but again more precisely
determined. Overall, the gluon fragmentation for both K$^*$ and $\phi$
have halved compared to the earlier analysis. This is because of
availability of $K^*$ and $\phi$ LHC data. Note that the same value of
$f_g^{\phi}$ is obtained without LHC data, on including the RHIC $p\,p$
data, since this is sufficient to capture sensitivity to the gluon
fragmentation. (RHIC does not have $K^*$ data and so the gluon
suppression factor for $K^*$ was earlier being determined only through
higher order effects in $e^+\,e^-$ hadroproduction. 

It is interesting that the obtained values of K$^*$ and $\phi$
gluon suppression parameters can be related as, $f_g^{\phi} \sim
(f_g^{K^*})^2$. This has an analogy with the assumption made in
the beginning that the sea suppression factor for $f_{sea}^{\phi} =
\lambda^2$, where $\lambda$ is the strangeness suppression in K$^*$. It
also explains the relatively large suppression of $\phi$ hadroproduction
in general: while non-strange fragmentation into $K^*$ is suppressed by
$\lambda$ since $K^*$ is a $q\overline{s}$ state, $q = u,d$, non-strange
fragmentation into $\phi$ is doubly suppressed by $\lambda^2$ since it
is dominantly a $s\overline{s}$ state. This appears to hold true for
both quark and gluon fragmentation. This may point to the presence of an
SU(3) symmetric sea of quarks and gluons.

\begin{figure}[htp]
\includegraphics[angle=-90,width=0.49\textwidth]{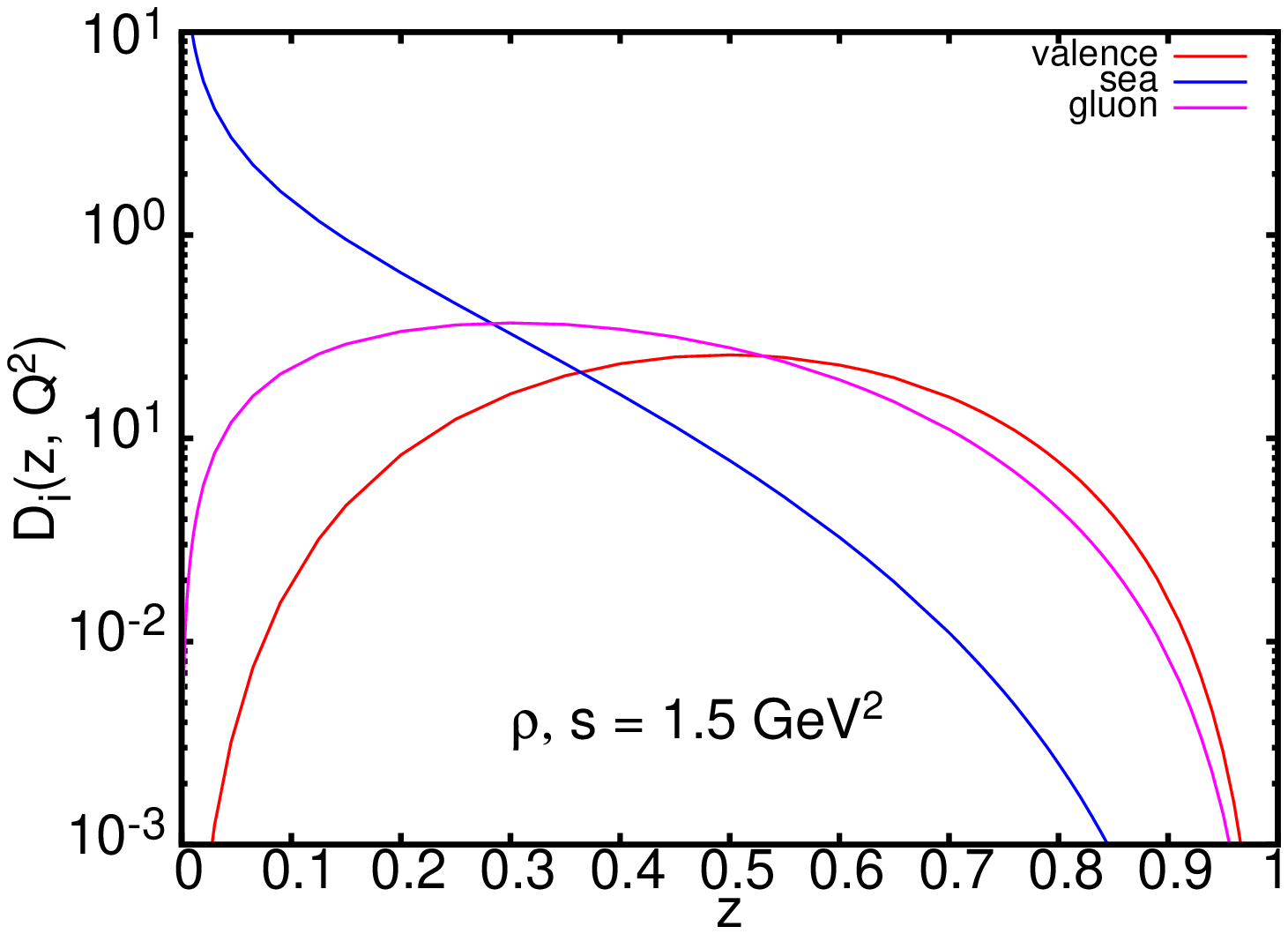}
\includegraphics[angle=-90,width=0.49\textwidth]{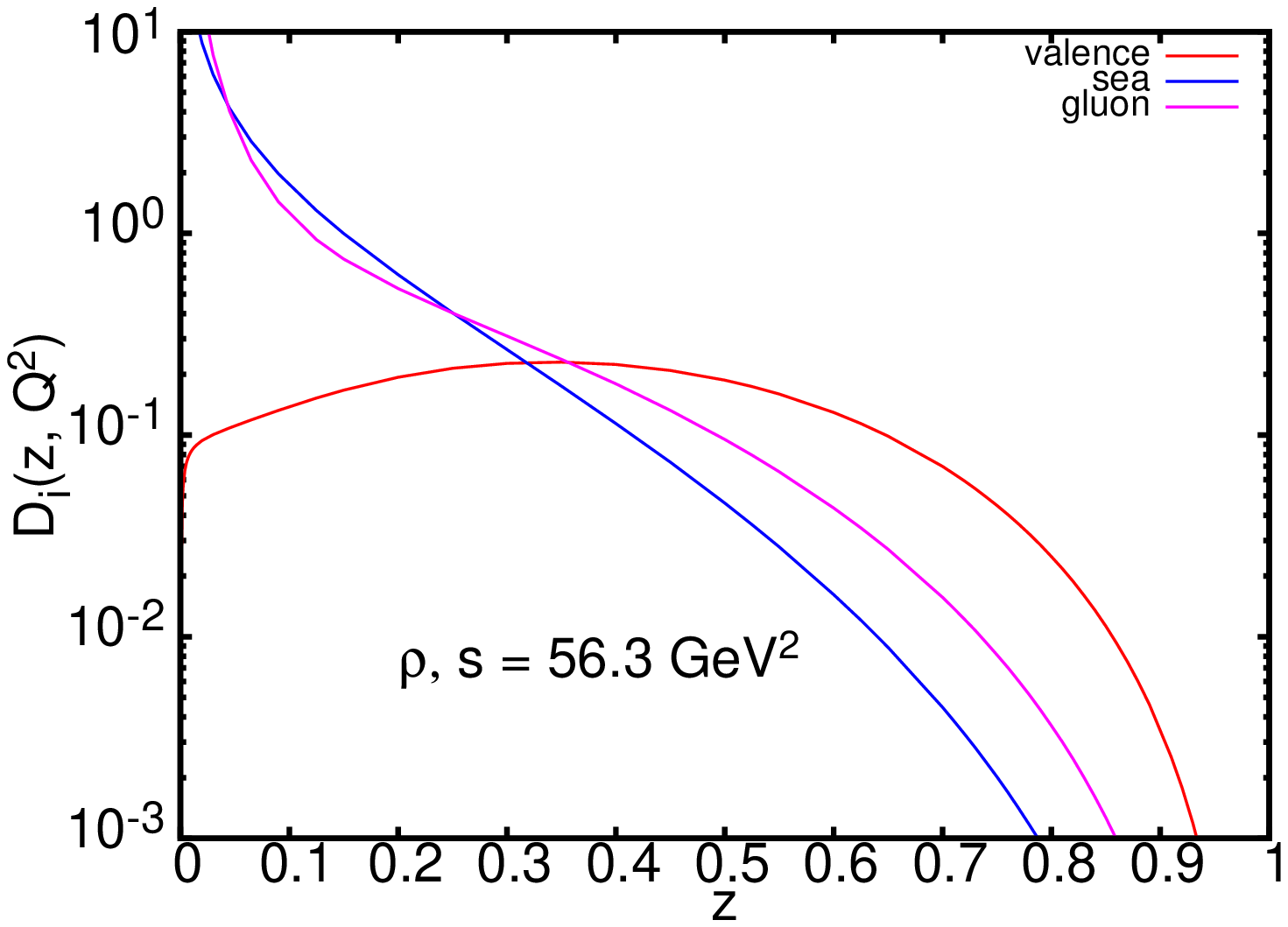}
\vspace*{8pt}
\caption{Initial fragmentation functions $D_i (z, Q^2$) at the starting
scale $Q^2 = Q_0^2 = 1.5$ GeV$^2$ (left) and fragmentation functions
at a sample value of $Q^2 = 56.3$ GeV$^2$ (right), $i=V, \gamma,D_g$,
for $\rho$ mesons as a function of the momentum fraction $z$.}
\label{fig:fragfns}
\end{figure}

The input fragmentation functions $(D_i(z,Q^2); i=$ valence, sea and
gluon) at a low energy scale of $Q_0^2 = 1.5$ GeV$^2$ for non-strange
$\rho$ mesons are plotted as a function of the momentum fraction ($z$)
as shown in the left of Fig.~\ref{fig:fragfns} for three light quarks
alone.  The figure on the right shows the fragmentation functions of the
same set for $\rho$ mesons at a sample value of $Q^2 = 56.3$ GeV$^2$,
where the heavy flavours are produced by gluon-initiated process through
evolution and not through heavy flavour B- or D- meson hadroproduction
and their subsequent decays.

\begin{figure}[htp]
\includegraphics[angle=-90,width=0.49\textwidth]{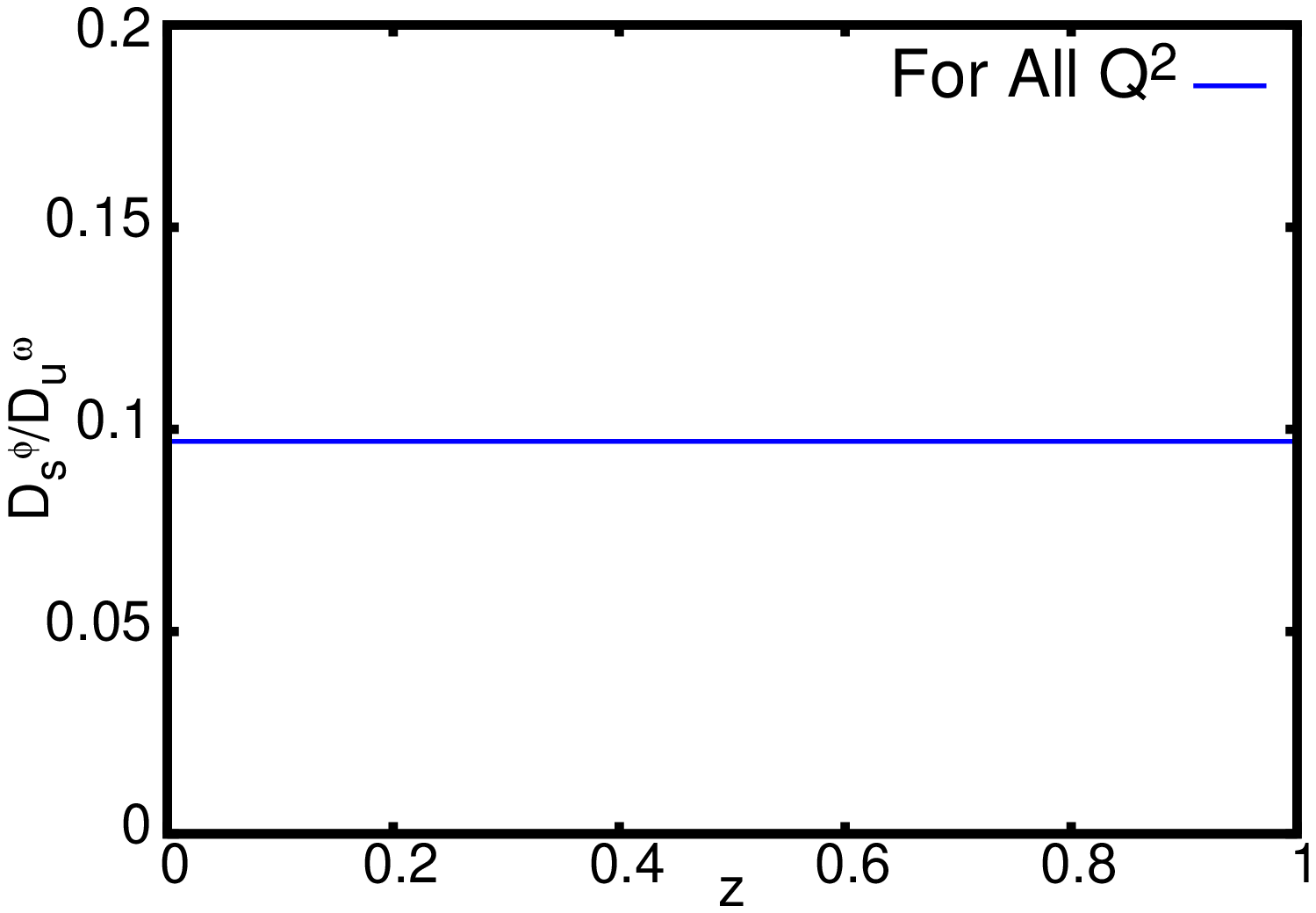}
\includegraphics[angle=-90,width=0.49\textwidth]{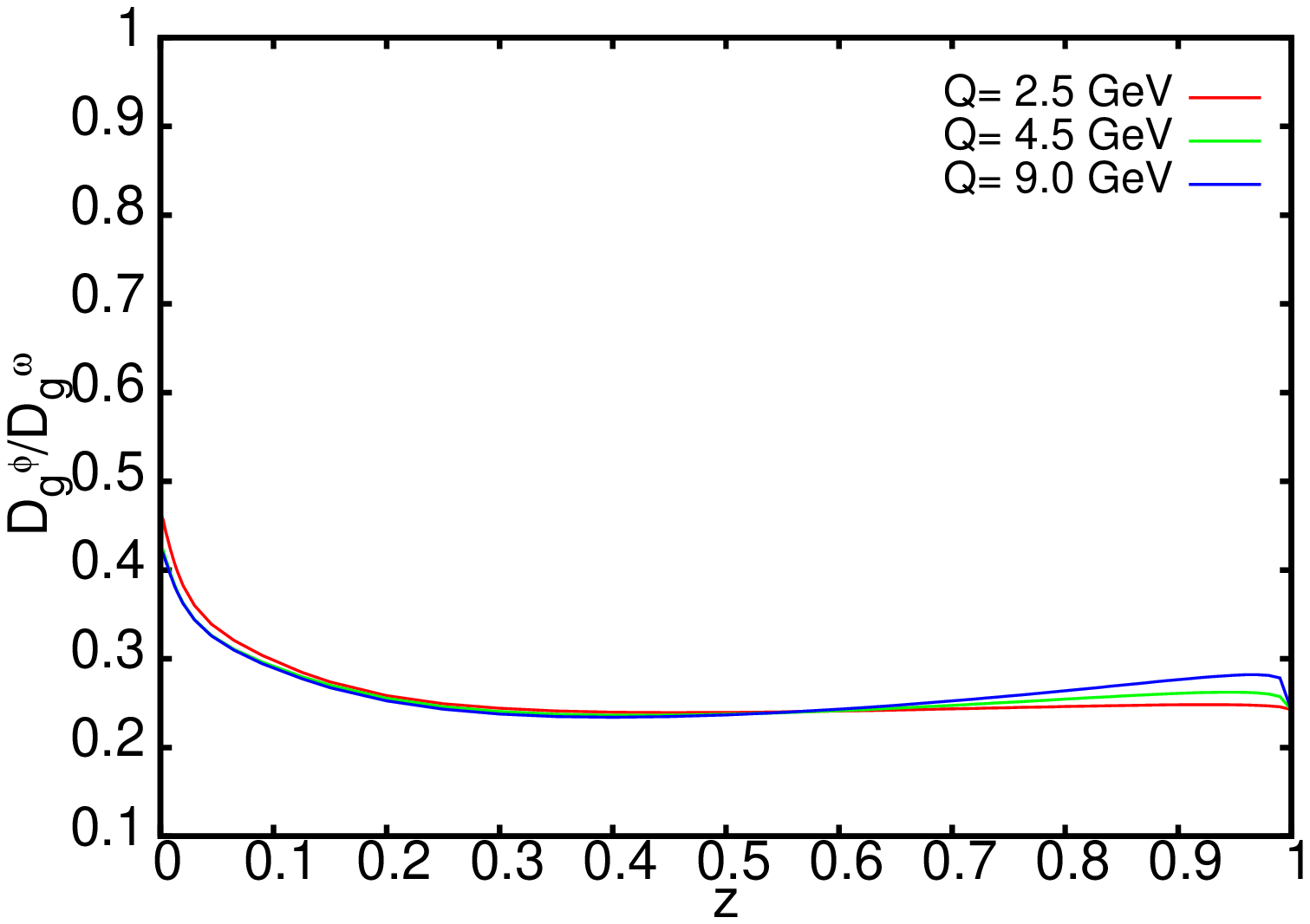}
\vspace*{8pt}
\caption{(Dominant) quark fragmentation function ratio (left) and gluon
fragmentation function ratio (right) of $\phi$ and $\omega$ mesons
as a function of the hadron momentum fraction $z$. While the quark
fragmentation function ratio is scale (and $x$-) independent, the gluon
fragmentation function ratio is shown for three different $Q^2 = p_T^2$
values.}
\label{fig:dqgratio}
\end{figure}

Fig.~\ref{fig:dqgratio} shows the ratio of the dominant quark
fragmentation functions (left) and gluon fragmentation functions (right)
for $\omega$ and $\phi$ mesons as a function of $x$. To understand the
quark ratio we considered the dominant non-strange quark fragmentation
functions $(D_u^{\omega})$ for $\omega$ mesons and strange quark
fragmentation functions $(D_s^{\phi})$ for $\phi$ mesons.  The quark
fragmentation ratio came out to be $D_s^{\phi}/D_u^{\omega} = \lambda$,
that is, equal to the strangeness suppression factor, $\lambda = 0.097$,
as expected, {\em independent} of $Q^2$. This shows that the model
captures strangeness suppression well at all values of $(x, Q^2)$.

The ratio of the gluon fragmentation functions for the two mesons came out
as $D_g^{\phi}/D_g^{\omega} \sim 0.25 \sim f_g^\phi/f_g^{\omega}$ for $x
> 0.1$,
as expected, whereas in the small-$x$ region $D_g^{\phi}/D_g^{\omega}$
rises towards
0.5, as can be seen from Fig.~\ref{fig:dqgratio}, hinting at SU(3)
symmetry restoration at small-$x$ for all $Q^2$. As far as we understand,
this interesting feature has not been pointed out before.

\subsection{Fits to $e^+\,e^-$ hadroproduction data}
\label{Datafits}
\begin{figure}[htp]
\includegraphics[angle=-90,width=0.49\textwidth]{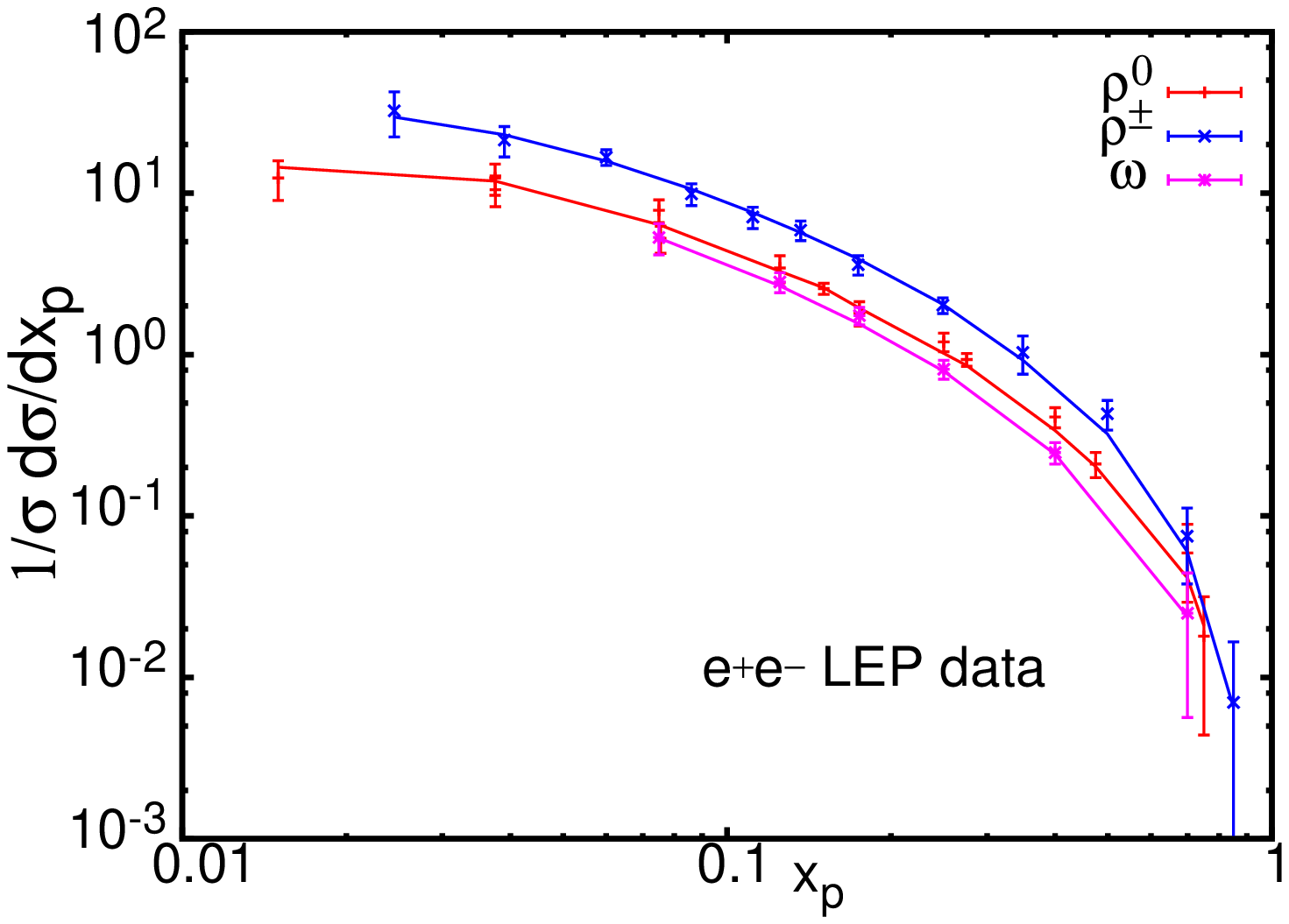}
\includegraphics[angle=-90,width=0.49\textwidth]{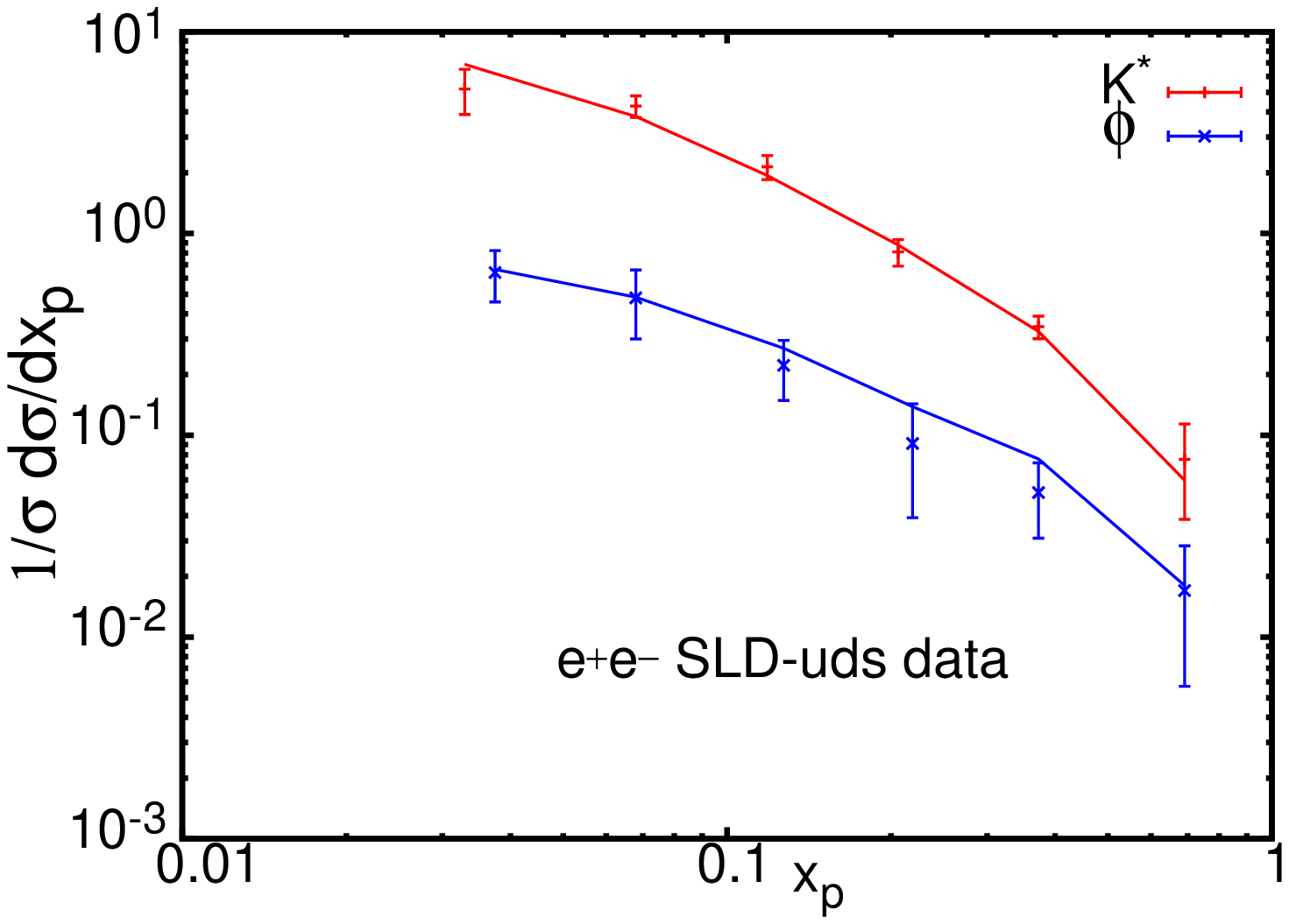}
\caption{Cross section behaviour as a function of $x_p$ for vector
meson fragmentation in $e^+\,e^-$ collisions. The data from
LEP\protect\cite{Data,Rho1,Rho2,Rho3,Rho5,Omega1,Omega2}
for $\rho^{\pm}, \rho^0$ and $\omega$, and ``pure uds data" for K$^*$
and $\phi$ mesons from SLD\protect\cite{SLD1,SLD2} at $\sqrt{s}
= 91.2$ GeV are shown in comparison with the solid lines which are the
best fits resulting from the present model.}
\label{fig:ee}
\end{figure}

The best fits to the input parameters obtained in the previous
section were used to evolve the fragmentation functions to the
$Z$-pole. The resulting cross-sections are shown in the left hand side
of Fig.~\ref{fig:ee} for $\rho^{\pm, 0}$ and $\omega^0$ hadroproduction
in comparison with the LEP
data\cite{Data,Rho1,Rho2,Rho3,Rho5,Omega1,Omega2} $e^+\,e^-$ data
and on the right for $K^{*,\pm,0,\overline{0}}$ and $\phi$ hadroproduction
in comparison with the SLD ``pure uds''\cite{SLD1,SLD2} $e^+\,e^-$ data.

There is good agreement with data throughout the $x$ range with an
overall $\chi^2 \sim 18.0 $ for 44 data points and 23 free parameters
with 21 degrees of freedom. For individual $\chi^2$ from each meson,
see Table~\ref{tab:chi2}.

\begin{table}[tbh]
\tbl{$\chi^2$ values obtained from best-fits to $\rho$, K$^*$, $\om$
and $\ph$ hadroproduction from $e^+\,e^-$ LEP, SLD data, and $\om$ and $\phi$ 
hadroproduction for central rapidity as well as ratio of branching fraction 
weighted cross sections of $\phi$ and ($\om+\rho$) mesons for forward rapidity from $p\,p$ RHIC-PHENIX data.}
{\begin{tabular}{ccccc} \toprule
Data Set         & No. of data points  &   $\chi^2$  \\  \colrule
Total {\emph{$e^+\,e^-$}} & 44 & 17.91 \\ \hline \hline
$\rho^0$         &   14                &     7.56     \\
$\rho^{+-}$      &   12                &     3.05     \\
K$^*$            &    6                &     3.65     \\
$\om$            &    6                &     1.02    \\
$\ph$            &    6                &     2.63     \\ \hline
Total {\emph{$p\,p$}(RHIC+LHC)} & 70 & 64.93 \\ \hline \hline
$\om$(RHIC)      &    33               &     16.89   \\
$\ph$(RHIC)      &    13               &     33.62    \\  \hline
$K^*$(LHC)       &    11               &     16.89   \\
$\ph$(LHC)       &    13               &     33.62    \\  \hline
Total            &    114               &    82.84    \\
{\emph{$e^+\,e^-$}}+ {\emph{$p\,p$}} & & \\ \hline
Total free parameters &    23           &     --    \\

Total {\emph{$e^+\,e^-$}+\emph{$p\,p$}} & $\chi^2$/dof & 82.84/91 \\ \hline \botrule
\end{tabular}
\label{tab:chi2}}
\end{table}

\subsection{Fits to $p\,p$ hadroproduction data}
\label{Ppfit}
The best-fit input parameters for the various fragmentation functions
obtained in the earlier section are evolved to different $p_T$ values to
obtain the differential hadroproduction cross-section defined in
Eq.~\ref{eq:physobs} in the central ($\vert y \vert \le 0.35$) and
forward ($1.2 \le y \le 2.2$) rapidity regions where data is available.
The best-fit (solid central) lines to the cross-section as a function of
$p_T$ is shown in Fig.~\ref{fig:band} for $\omega$ (left) and $\phi$
(right) in comparison with central rapidity data from RHIC-PHENIX\cite{RHIC}.

The fits reflect the consistency and power of the model which explains two
entirely different processes without introduction of any new parameter in
the analysis.  We have analysed both the momentum ($p_T$) and rapidity
($y$) dependence of differential cross section for $\omega$ and $\phi$
mesons.

\subsubsection{Scale ($p_T$) dependence} 
\label{sssec:num1}

\begin{figure}[htp]
\includegraphics[angle=-90,width=0.49\textwidth]{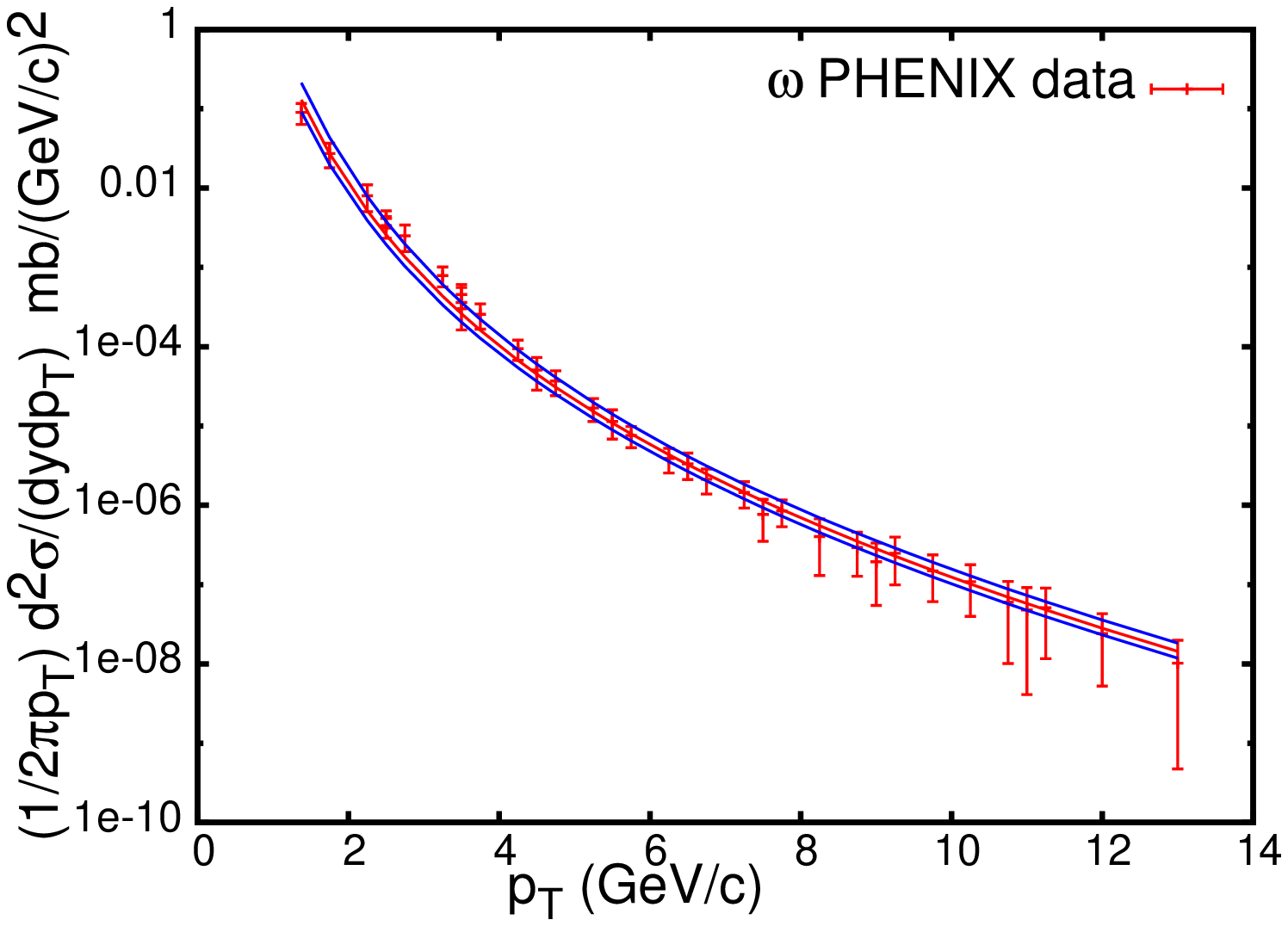}
\includegraphics[angle=-90,width=0.49\textwidth]{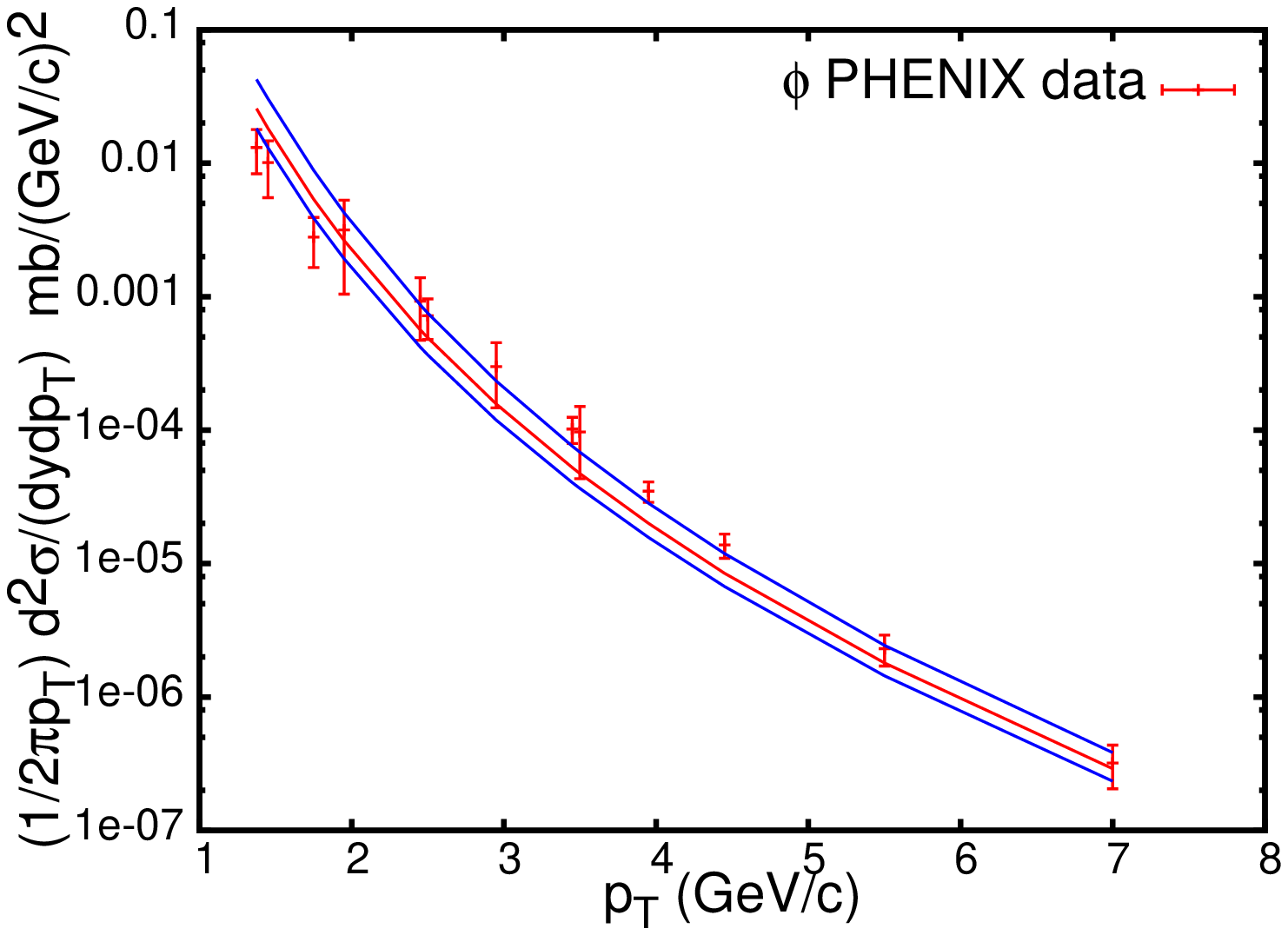}
\vspace*{8pt}
\caption{Cross section as a function of $p_T$ for $\omega$ (L) and $\phi$
(R) meson hadroproduction in $p\,p$ collisions at $\sqrt{s} = 200$ GeV and
$\vert y\vert \ \le 0.35$. Bands show the scale uncertainty on changing
$Q^2=p_T^2$ over a range $p_T^2/2$ (upper curve) $\le Q^2 \le 2 p_T^2$
(lower curve) for all the three scales. See text for more details.}
\label{fig:band}
\end{figure}

In the $p\,p$ hadroproduction process, the factorization, renormalization
and the fragmentation scales are made equal to the transverse momentum,
$M \sim \mu \sim M_f \sim p_T$ and the uncertainty in the scales are
determined by changing the value of $Q^2 = p_T^2$ over a range $p_T^2/2
\leq Q^2 \leq 2p_T^2$.

Fig.~\ref{fig:band} clearly shows the effect of the scale uncertainty for
both the mesons. Keeping all the three scales $(M, \mu, M_f)$ equal to
$Q$ for convenience and changing $Q^2$ from $p_T^2/2$ to $2p_T^2$ gives
an uncertainty band as shown in the figure.  The central curve in both
left and right side of the Fig.~\ref{fig:band} is the actual fit without
scaling for original $Q^2=p_T^2$, whereas the upper curve is for a scale
change of $Q^2 = (1/2) p_T^2$ for all the three scales and the lower curve
for $Q^2 = 2 p_T^2$. For the $\om$ meson, the scale uncertainty is seen
to be rather small, whereas for the $\phi$ meson the scale dependence
is significant. The reduced scale uncertainty for mesons at NLO compared
to the earlier LO analysis\cite{Savilo} shows that the scale dependence
decreases with inclusion of higher order terms, as is expected.

\begin{figure}[htp]
\includegraphics[angle=-90,width=0.49\textwidth]{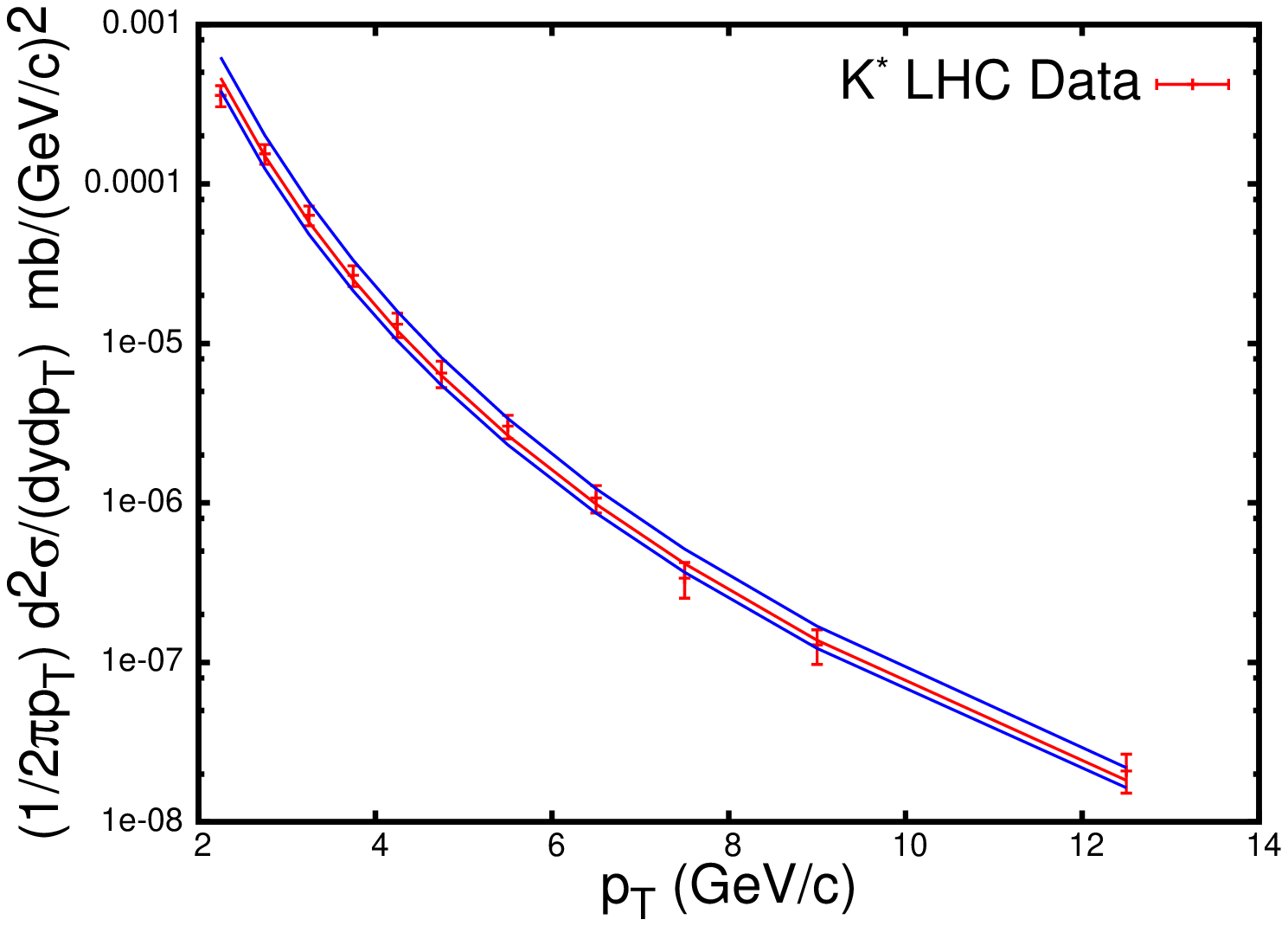}
\includegraphics[angle=-90,width=0.49\textwidth]{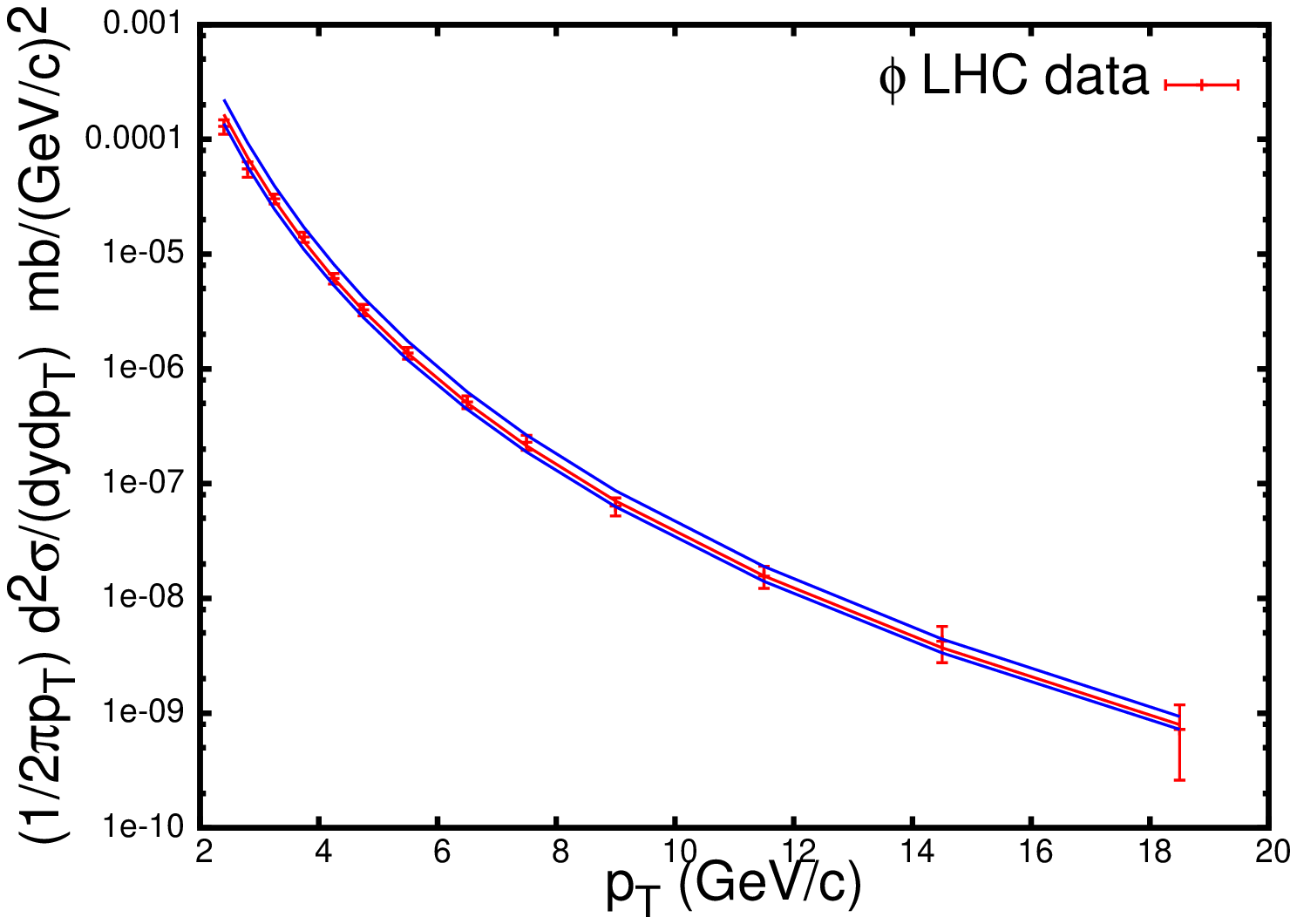}
\vspace*{8pt}
\caption{Cross section as a function of $p_T$ for K$^*$ (L) and $\phi$
(R) meson hadroproduction in $p\,p$ collisions at $\sqrt{s} =2.76$ TeV and
$\vert y\vert \ \le 0.5$. Bands show the scale uncertainty on changing
$Q^2=p_T^2$ over a range $p_T^2/2$ (upper curve) $\le Q^2 \le 2 p_T^2$
(lower curve) for all the three scales.}
\label{fig:lhcband}
\end{figure}

The scale dependence is even smaller for the LHC data as can be seen
from Fig.~\ref{fig:lhcband}. This is the main reason for the small
errors in our fits.

\subsubsection{Fits to the rapidity dependence of $p\,p$ data}
\label{sssec:num3}

The RHIC-PHENIX collaboration has also studied\cite{Ratio} the branching
ratio (BR)- weighted differential cross section of $(\rho+\om)$ and
$\phi$ as a function of rapidity over the $p_T$ range from $1 \le p_T
\le 7$ GeV$/c$. The event rates are defined as
\begin{eqnarray} \nonumber
(N_{\om}+ N_{\rho}) & = & \left(BR(\om\rightarrow \mu\mu)\sigma_{\om} +
 BR(\rho\rightarrow \mu\mu)\sigma_{\rho}\right)~, \\
N_\phi & = & BR(\phi\rightarrow \mu\mu)\sigma_{\phi}~, \nonumber
\end{eqnarray}
where the relevant branching ratio to dimuons for $\rho$ is $(4.55\pm0.28)
\times 10^{-5}$, for $\om$ is $(9.0\pm3.1)\times 10^{-5}$ and for $\phi$
is $(2.87\pm 0.19) \times 10^{-4}$ as given by both
the RHIC-PHENIX collaboration\cite{Ratio} and by the Particle Data Group\cite{PDG}.

\begin{figure}[hpt]
\includegraphics[angle=-90,width=0.49\textwidth]{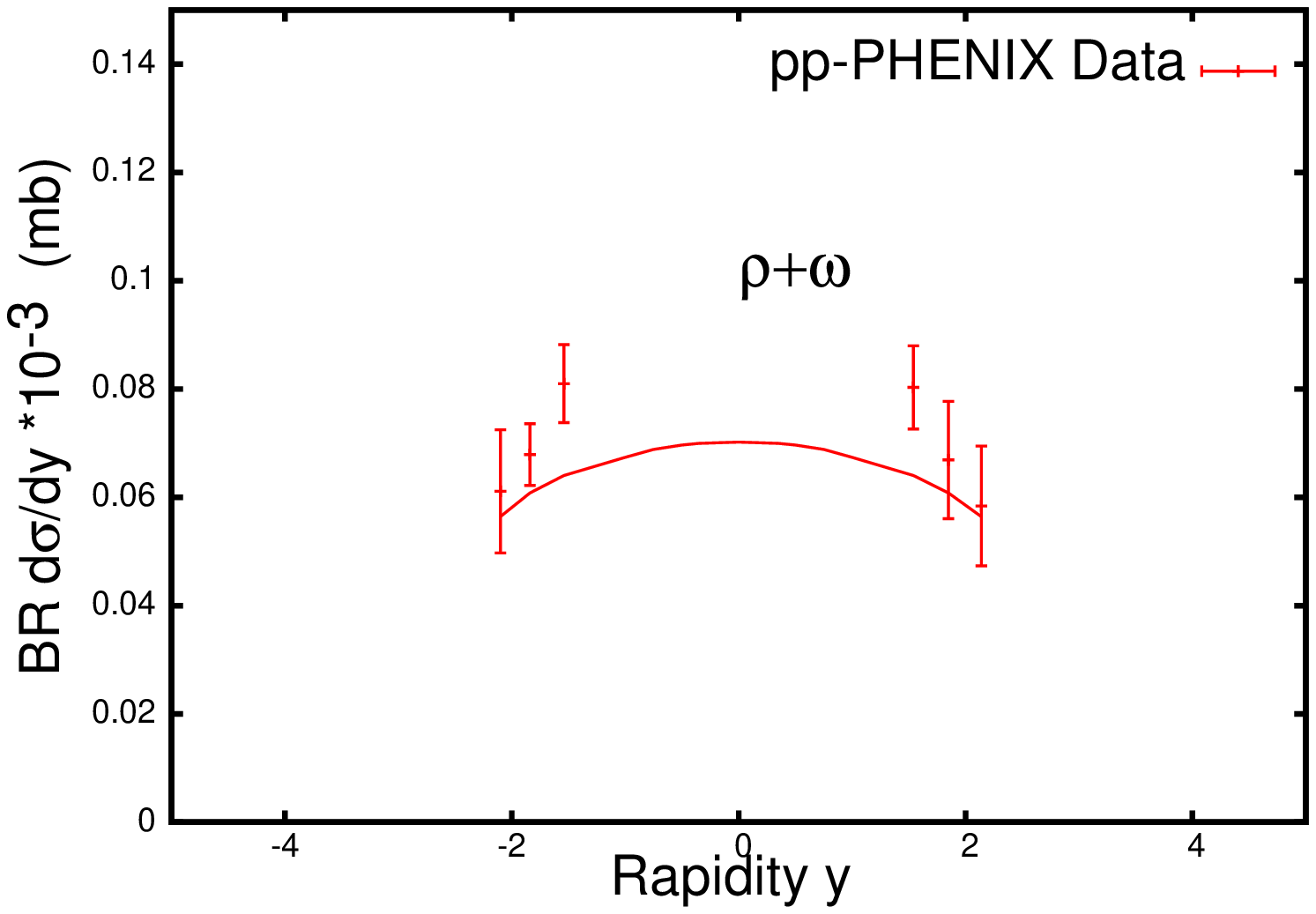}
\includegraphics[angle=-90,width=0.49\textwidth]{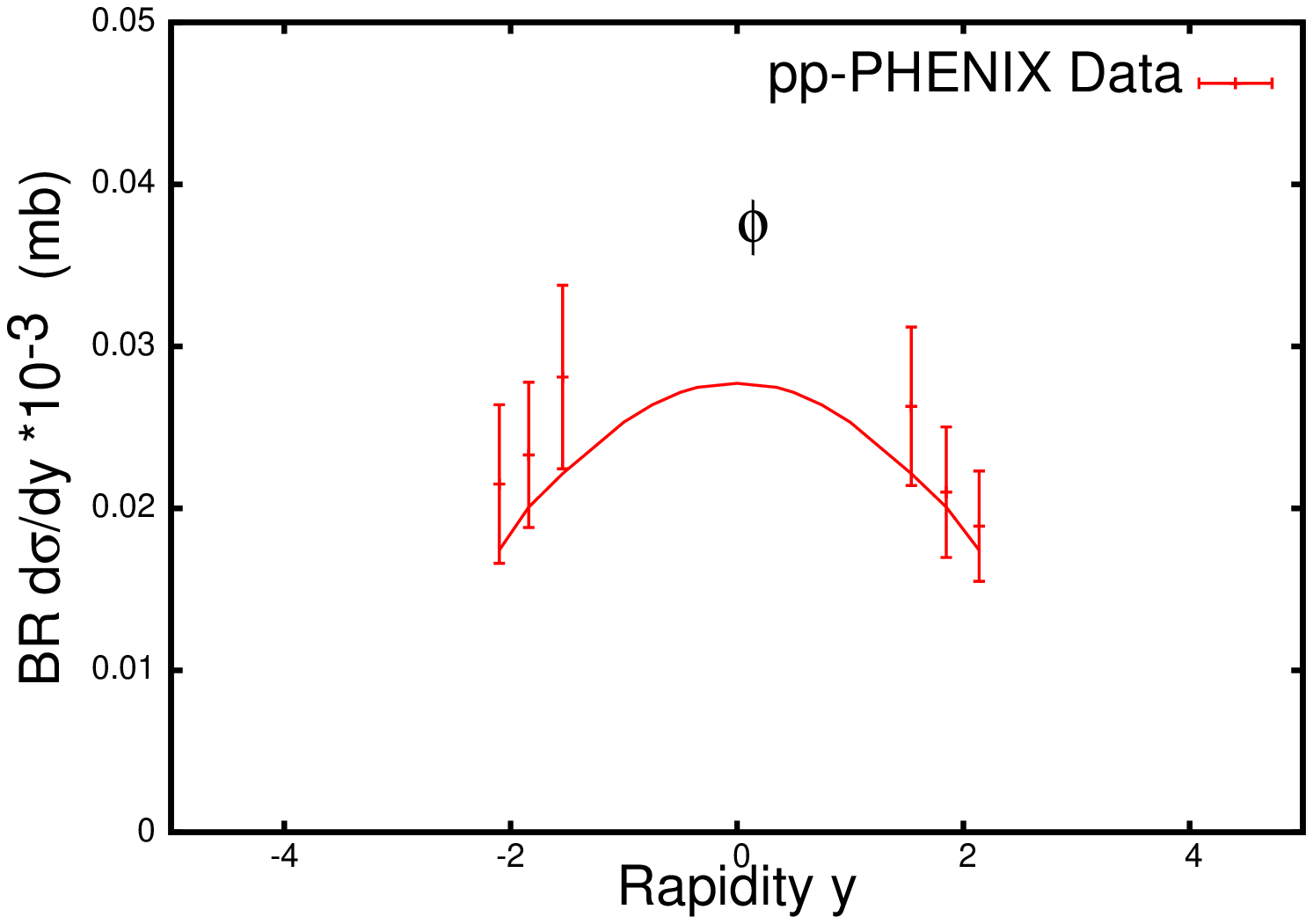}
\vspace*{8pt}
\caption{Solid lines show the model fit to the branching fraction-weighted
differential cross sections as a function of rapidity, $y$, for
$(\rho+\omega)$ (L) and $\phi$ (R) meson hadro-production in $p\,p$
collisions at $\sqrt{s} = 200$ GeV in comparison with the RHIC data. Both
statistical $\&$ systematical errors are added in quadrature.}
\label{fig:rapidity}
\end{figure}

Here $\sigma_i$ is the integrated cross-section, $\sigma_i =
d\sigma/dy$, $i = \omega, \rho, \phi$, and the model calculation has
been performed by integrating from $1.225 \le p_T (\hbox{GeV}) \le 7$
since the starting scale is $Q_0^2 = 1.5$ GeV$^2$.

Fig.~\ref{fig:rapidity} shows that the cross-section for hadroproduction
of non-strange mesons
like $\rho$ and $\omega$ fall slower
with rapidity from central to forward regions and are barely consistent
with the data (see Table~\ref{tab:rapidity}) while
in the case of $\phi$ hadroproduction, the model fits well
with the data. While the fits are still reasonably good, an improvement
in the error bars of the data will severely constrain the model
parameters, especially that of the gluon fragmentation functions. In
fact, reproducing this rapidity dependence was the biggest constraint in
determining the model parameter fits.

\begin{table}[hpb]
\tbl{Differential cross sections for hadroproduction in $p\,p$
collisions, weighted by branching fractions, as a function
of rapidity obtained from model best-fits for $\rho+\om$ and $\phi$
mesons for $1.22 \le {p_T} \le 7$ GeV/c and $\sqrt{s} = 200$ GeV.}
{\begin{tabular}{ccccc} \toprule
$y$    & \multicolumn{2}{c}{$\frac{(BR d\s)_{\rho+\om}}{dy}$ (nb)}&
\multicolumn{2}{c}{$BR{d\s_{\phi}}/dy$ (nb) }     \\
\cline{2-3}
\cline{4-5}
     &    fit  &  Data                 &   fit   &   Data            \\ \colrule
-2.14 &  56.44  & 61.1$\pm$6.7$\pm$9.2  &  17.42  & 21.5$\pm$3.7$\pm$3.2 \\
-1.85 &  60.82  & 67.9$\pm$5.6$\pm$10.2 &  20.08  & 23.3$\pm$2.8$\pm$3.5 \\
-1.54 &  64.04  & 81.0$\pm$7.1$\pm$12.2 &  22.15  & 28.1$\pm$3.8$\pm$4.2 \\
1.54  &  64.04  & 80.3$\pm$7.6$\pm$11.2 &  22.15  & 26.3$\pm$3.2$\pm$3.7 \\
1.85  &  60.82  & 66.9$\pm$5.4$\pm$9.4  &  20.08  & 21.0$\pm$2.8$\pm$2.9 \\
2.14  &  56.44  & 58.4$\pm$7.4$\pm$8.2  & 17.42   & 18.9$\pm$2.2$\pm$2.6\\\botrule
\end{tabular}
\label{tab:rapidity}}
\end{table}

\subsubsection{Events ratio} 
\label{sssec:num4}

The event ratio is given by,
$$
\frac{N_{\phi}}{(N_{\om}+ N_{\rho})}= \frac{BR(\phi\rightarrow \mu\mu)\sigma_{\phi}}
{(BR(\om\rightarrow \mu\mu)\sigma_{\om} + BR(\rho\rightarrow
\mu\mu)\sigma_{\rho})}~,
$$
where the ratio is determined for $1.22 \le p_T \le 7$ GeV/c,
for both central ($\vert y \vert \le 0.35$) and forward rapidity ($1.2
\le \vert y\vert  \le 2.2$) regions.

\begin{figure}[htp]
\begin{center}
\includegraphics[angle=-90,width=0.49\textwidth]{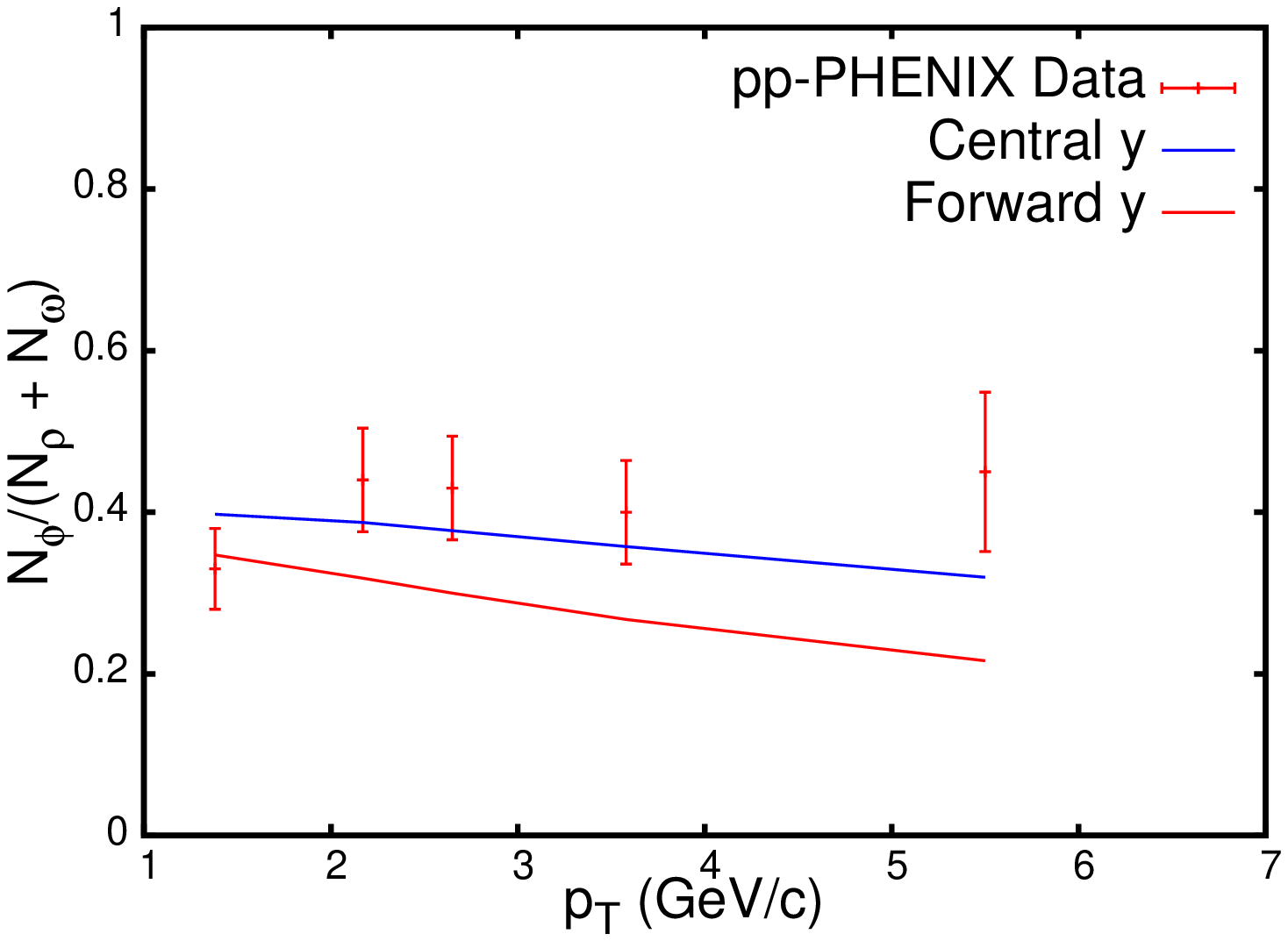}
\caption{Model best-fits to $N_{\phi}/(N_{\om}+N_{\rho})$ as a function
of $p_T$ over the range $1.22 \le p_T \le 7$ GeV$/c$ for $\sqrt{s} =
200$ GeV in comparison with the RHIC data\protect\cite{Ratio}. The statistical
and systematical errors are added in quadrature. The upper solid line
represents the fit for the central rapidity region ($\vert y \vert  \le
0.35$) while the lower one is for the forward rapidity region ($1.2 \le
\vert y \vert \le 2.2$).}
\label{fig:ratio}
\end{center}
\end{figure}

\begin{table}[hpb]
\tbl{The ratio $N_{\phi}/(N_{\rho}+N_{\om})$ vs. $p_T$ in (GeV/c)
for both central ($\vert y \vert  \le 0.35$) and forward ($1.2 \le \vert
y\vert \le 2.2$) rapidity regions for $\sqrt{s} = 200$ GeV. For more
details, see text.}
{\begin{tabular}{ccccc} \toprule
$p_T$    &  \multicolumn{3}{c}{$N_{\phi}/(N_{\rho}+N_{\om})$} \\
\cline{2-4}
(GeV/c)  & Central $y$ & Forward $y$ & Data for forward $y$    \\ \colrule
 1.375   &   0.398    &   0.347     &  0.33$\pm$0.04$\pm$0.03 \\
 2.2     &   0.387    &   0.319     &  0.44$\pm$0.05$\pm$0.04 \\
 2.65    &   0.377    &   0.299     &  0.43$\pm$0.05$\pm$0.04 \\
 3.5     &   0.358    &   0.267     &  0.40$\pm$0.05$\pm$0.04 \\
 5.5     &   0.320    &   0.216     &  0.45$\pm$0.09$\pm$0.04 \\ \botrule
\end{tabular}
\label{tab:ratio}}
\end{table}

The model values for the ratio $N_{\phi}/(N_{\om} + N_{\rho})$ are
listed in Table~\ref{tab:ratio} for central as well as forward rapidity
regions. The ratio was determined as $0.40$ in the central region,
on the average, whereas it was found to be $0.30$ in the forward
rapidity region. The ratio for both central and forward rapidity
regions are in agreement with the data value of $0.390 \pm 0.021$
(stat) $\pm 0.035$ (sys) since the data have large statistical and
systematical uncertainitites. However, the detailed $p_T$ dependence
is not correctly reproduced, especially for forward rapidities, as can
be seen from Fig.~\ref{fig:ratio}. This is a reflection of the slower
fall with increasing rapidity $\vert y \vert$ in $(\rho + \omega)$ as
discussed earlier and shown in Fig.~\ref{fig:rapidity}. With more data,
presumably, the fits in this sector can be improved in the future.

The $\chi^2$ values obtained from fits to K$^*$, $\om$ and $\phi$ mesons
in the central rapidity region and the branching fraction-weighted
ratios in the forward rapidity region of $(\rho+\omega)/\phi$ are given in Table~\ref{tab:chi2}. 
The model provides reasonable fits to the parameters with reduced error 
bars with a $\chi^2 = 65$ for 70 data points excluding the ratio.
Apart from
individual values, an overall $\chi^2$ of $83$ is obtained from the
combined $e^+\,e^-$ and $p\,p$ (hadroproduction in the central region
excluding LHC data with $p_T < 2$ GeV) data with 114 data points, 23 fit
parameters and hence  $91$ degrees of freedom,  which is pretty 
good, and reflects the consistency and efficacy of the model.

An effort was made to understand the LHC/ALICE data\cite{LHC} for the
production of $K^*$ and $\ph$ mesons in $p\,p$ collisions
and also the branching fraction weighted ratios at $\sqrt{s} = 7$ TeV.
The data has $p_T$ values ranges from $1.25$ to $5.5$
GeV. Therefore, at these low $p_T$, the $x$-values will be of the
order of $10^{-6}$ and parton distribution functions are not available
at such low values of $x$. In addition, the $z$ values are less than $z
< 10^{-3}$. It is well known that DGLAP evolution equations for
fragmentation functions
fail at such small $z$-values due to the poles in both the $P_{gq}$
and $P_{gg}$ splitting functions which cause both the singlet quark and
gluon fragmentation functions to diverge at small-$z$. Such studies at low
$z$-values can be done using modified leading log approximation
(MLLA)\cite{Indu} which yield better (convergent) behaviour of fragmentation
functions at small $z$. This is beyond the scope of the present work.

\section{Conclusion}
\label{Sum}
Vector meson fragmentation has been studied for the first time in both
$e^+\,e^-$ and $p\,p$ collisions at NLO with the comparison of LEP,
SLD ($e^+\,e^-$) and RHIC ($p\,p$) data using a model with broken SU(3)
symmetry. While this work was being completed the authors became aware
of $p\,p$ hadroproduction data from LHC-ALICE which tremendously
improved the fit values.

The model with three light flavours $u, d$ and $s$ uses SU(3) symmetry
to describe the unknown fragmentation functions in terms of three
independent quark fragmentation functions $\alpha(x, Q^2)$, $\beta(x,
Q^2)$ and $\gamma(x, Q^2)$ with their conjugates and a gluon fragmentation
function. The model uses further symmetries like isospin invariance and
charge conjugation to reduce the functions to two universal functions,
the valence $V(x, Q^2)$ and sea $\gamma(x, Q^2)$ quark fragmentation
functions and a gluon fragmentation function $D_g(x, Q^2)$.

A strangeness suppression parameter $\lambda$ describes strangeness
suppression in $K^*$ mesons. The entire meson nonet (and hence the
physical $\omega$ and $\phi$ hadroproduction) is considered by including
a singlet--octet mixing parameter, $\theta$. Instead of introducing a (yet
another unknown) singlet fragmentation function, this is related to the
octet fragmentation function $\alpha(x, Q^2)$ through two
proportionality constants, one each for $\omega$ and $\phi$ mesons along
with a parameter that describes strangeness suppression in $\omega$
(which turns out to be $\sim 1$ since $\omega$ is dominantly a
non-strange meson due to the particular value of the mixing angle). The
strangeness suppression factor in $\phi$ turned out to be close to
$\lambda^2$, albeit with larger error bars, and was set to be equal to
$\lambda^2$. No new fragmentation function or additional parameters are
introduced in order to explain the $p\,p$ hadroproduction data. Finally,
individual gluon suppression factors were introduced for $\omega$,
$K^*$, and $\phi$, although the first was close to unity.

The best-fit values of the 23 free parameters are given in Tables
\ref{tab:inputs} and \ref{tab:unknown_par}.

The new gluon dependent parameter values are determined more precisely
with reduced error bars (within $5\%$) compared to the previous analysis
with $e^+\,e^-$ data alone\cite{Savinlo}.  The K$^*$ and $\phi$ gluon
suppression values are related by $f_g^{\phi}\sim (f_g^{K^*})^2$ similar
to the result, $f_{sea}^\phi = (f_{sea}^{K*})^2$; this can be used to
further reduce the number of fit parameters. This shows the stability of
the model and indicates the presence of an SU(3) symmetric sea of quarks
and gluons over the entire nonet. Furthermore, the ratio of (dominant)
quark fragmentation for $\phi$ and $\omega$ mesons came out to be equal
to $\lambda=0.097$, the strangeness suppression parameter, which implies
that $\omega$ is dominantly a non-strange meson and $\phi$ is dominantly
a $s\overline{s}$ state. In contrast the corresponding gluon ratio tended
to rise at low $x$, indicating restoration of SU(3) symmetry here.

The model explains both the $e^+\,e^-$ and $p\,p$ data with good
fits and reasonable $\chi^2$ as seen from Table \ref{tab:chi2} and
Figs.~\ref{fig:ee} and \ref{fig:band}.

The $p_T$ band in Fig. \ref{fig:band} shows the reduced scale
dependence for $p\,p$ hadroproduction at NLO compared to earlier
results\cite{Savilo} at LO, as expected. This is even smaller for the LHC
data. The branching ratio-weighted differential
cross section for $\rho+\omega$ and $\phi$ hadroproduction in $p\,p$
collisions as a function of rapidity were fitted with the RHIC
data\cite{Ratio}. The results show that the rates for $\rho$ and
$\omega$ mesons fall slower with increase in rapidity away from the
central region whereas for $\phi$ it is comparatively faster; while the
fits to $\phi$ measons are consistent with data, that for $\omega$ are
barely consistent with the data.

Finally, the ratio of branching fraction-weighted cross-sections for
$\phi$ and $(\omega+\rho)$ mesons was found to be 0.40
for central rapidity and 0.30 for forward rapidity regions; this agrees
with the data while not being fully in agreement with the detailed $p_T$
dependence of the data. Note that the low $p_T$ region is particularly
intractable because of large changes upon even very small evolution from
the starting scale $Q_0^2 = 1.5$ GeV$^2$. It may be possible to tune the
parameter fits to improve the agreement in this sector with the
availability of more data with improved error bars.

This completes the program of describing the fragmentation functions of
the entire vector meson nonet at NLO using both $e^+\,e^-$ and $p\,p$
hadroproduction data. The fragmentation of vector mesons like $\om$
and $\phi$ meson which have been studied for $p\,p$ collisions with RHIC
data will be useful to understand, for example, strangeness suppression
or $\phi$ production in nucleus-nucleus collisions as a signal in QGP
studies. This was one of the primary motivations for this work. A
table of quark and gluon fragmentation functions for all vector mesons
is available at the web-site of one of the authors\cite{web}. A sample
fortran code that can be used to generate the fragmentation functions
at any $(Q^2, x)$ using linear interpolation is also available.

\section*{Acknowledgement}
\label{Acknow}
One of the authors HS would like to acknowledge the financial support
in the form of fellowship from High Energy Physics Project of Institute
of Mathematical Sciences, Chennai and M V N Murthy for suggestions. The
authors thank the referee for valuable suggestions and for pointing us
to the LHC hadroproduction data.

\end{document}